\documentclass[aps,prb,twocolumn,superscriptaddress]{revtex4-1}

\usepackage{graphicx}
\usepackage{amsmath}
\usepackage{mathtools}
\usepackage{color}

\begin{document}


\title{Practical $GW$ scheme for electronic structure of 3$d$-transition-metal monoxide anions: ScO$^{-}$, TiO$^{-}$, CuO$^{-}$, and ZnO$^{-}$} 



\author{Young-Moo Byun}
\affiliation{Department of Physics, University of Illinois at Chicago, Chicago, IL 60607, USA}

\author{Serdar \"{O}\u{g}\"{u}t}
\email[]{ogut@uic.edu}
\affiliation{Department of Physics, University of Illinois at Chicago, Chicago, IL 60607, USA}


\date{\today}

\begin{abstract}
The $GW$ approximation to many-body perturbation theory is a reliable tool for describing charged electronic excitations, and it has been successfully applied to a wide range of extended systems for several decades using a plane-wave basis. However, the $GW$ approximation has been used to test limited spectral properties of a limited set of finite systems (e.g. frontier orbital energies of closed-shell $sp$ molecules) only for about a decade using a local-orbital basis. Here, we calculate the quasiparticle spectra of closed- and open-shell molecular anions with partially and completely filled 3$d$ shells (shallow and deep 3$d$ states, respectively), ScO$^{-}$, TiO$^{-}$, CuO$^{-}$, and ZnO$^{-}$, using various levels of $GW$ theory, and compare them to experiments to evaluate the performance of the $GW$ approximation on the electronic structure of small molecules containing 3$d$ transition metals. We find that the $G$-only eigenvalue self-consistent $GW$ scheme with $W$ fixed to the PBE level ($G_{n}W_{0}@\text{PBE}$), which gives the best compromise between accuracy and efficiency for solids, also gives good results for both localized ($d$) and delocalized ($sp$) states of 3$d$-transition-metal oxide molecules. The success of $G_{n}W_{0}@\text{PBE}$ in predicting electronic excitations in these systems reasonably well is likely due to the fortuitous cancellation effect between the overscreening of the Coulomb interaction by PBE and the underscreening by the neglect of vertex corrections. Together with the absence of the self-consistent field convergence error (e.g. spin contamination in open-shell systems) and the $GW$ multi-solution issue, the $G_{n}W_{0}@\text{PBE}$ scheme gives the possibility to predict the electronic structure of complex real systems (e.g. molecule-solid and $sp$-$d$ hybrid systems) accurately and efficiently.
\end{abstract}



\maketitle 


\section{Introduction} \label{sec:introduction}

It is a challenging task to accurately determine the electronic structure of an interacting many-electron system. In experiment, electron removal and addition energies of both extended and finite systems are measured by direct and inverse photoelectron spectroscopy (PES and IPES, respectively). In theory, it is well known that the $GW$ approximation to many-body perturbation theory (MBPT) describes bandgaps and band structures of solids more accurately than local and semi-local approximations to density-functional theory (DFT).~\cite{Bruneval14,Reining18} However, less is known about the performance of the $GW$ approximation on the electronic structure of atoms, molecules, and clusters. Especially, $GW$ calculations for the quasiparticle (QP) spectra of open-shell molecules containing 3$d$ transition metals are scarce. There are a few reasons for it.

First, it is easier to test only frontier orbital energies such as the ionization energy (IE) and the electron affinity (EA) than the full QP spectrum (all orbital energies). There are mainly two ways to calculate IE and EA of molecules. On one hand, IE (EA) can be obtained from DFT, HF (Hartree-Fock), MP2 (second-order M{\o}ller-Plesset perturbation theory), RPA (random-phase approximation), or CCSD(T) (coupled-cluster singles and doubles plus perturbative triples) total energy differences between a neutral and a cation (anion) within the so-called $\Delta$SCF (self-consistent field) method.~\cite{Ren12} Generally, the $\Delta$SCF method gives good results for frontier orbital energies of finite systems, but cannot be applied to extended systems. Also, it is not straightforward for the $\Delta$SCF method to access the full QP spectrum. On the other hand, IE and EA can be obtained from $GW$ eigenvalues for the HOMO (highest occupied molecular orbital) and the LUMO (lowest occupied molecular orbital), respectively. Due to the simplicity of the $\Delta$SCF method, many studies have evaluated the performance of the $GW$ approximation on molecules by comparing $GW$ IE and EA to $\Delta$SCF ones,~\cite{Bruneval13} but that approach does not utilize the full power of the $GW$ approximation, which is the ability to provide the QP spectrum for both finite and extended systems. For example, Bethe-Salpeter equation (BSE) calculations for optical excitations require more orbital energies than IE and EA as input.~\cite{Pavarini16}


Second, it is easier to test closed-shell systems than open-shell ones. Most of quantum chemistry-based $GW$ implementations for finite systems, such as MOLGW,~\cite{Bruneval16} FIESTA,~\cite{Blase11} TURBOMOLE,~\cite{vanSetten13} FHI-AIMS,~\cite{Ren12} and CP2K,~\cite{Wilhelm16} use local-orbital basis sets such as Gaussian basis sets. $GW$ calculations require mean-field self-consistent calculations, such as restricted and unresticted Hartree-Fock or Kohn-Sham (RHF or RKS and UHF or UKS, respectively) calculations for closed- and open-shell systems, respectively. The problem is that unlike RHF and RKS self-consistent calculations, UHF and UKS ones are not guaranteed to converge, as their convergence strongly depends on the initial guess wavefunctions. This is especially the case for spin-unrestricted calculations performed with hybrid exchange-correlation (xc) functionals, which include a fraction of exact exchange (EXX), and HF on 3$d$-transition-metal-containing molecules due to the near-degeneracy of energy levels.~\cite{Gutsev00,Gonzales00,Kudin07} Partially due to this SCF convergence issue, most existing studies have used only closed-shell systems to assess the performance of the $GW$ approximation on finite systems. For example, Refs.~\onlinecite{vanSetten15,Caruso16,Maggio17,Govoni18} used the so-called $GW$100 benchmark set, which is composed of only closed-shell molecules.

Last, it is easier to test $sp$-electron systems than $d$-electron ones. Fundamentally, it is more difficult to accurately predict the electronic structure of $d$ systems (especially, 3$d$ systems) than $sp$ ones because of the strong localization, and thus the strong correlation, of $d$ electrons. For example, it is challenging for $GW$ to accurately reproduce the experimental bandgap and $d$-band position of bulk ZnO at the same time.~\cite{Shishkin07,Fuchs07,Klimes14} Practically, it is computationally more demanding to tackle systems with $d$-electrons than those with only $sp$-electrons. For example, $d$ elements have more basis functions than $sp$ ones, which increases the computational effort, and transition-metal-containing molecules, especially with partially filled $d$ shells and low multiplicity states, aggravate the above-mentioned SCF convergence issue, which increases the human effort by making it necessary to manually explore many minima with similar energies using many initial guess wavefunctions.~\cite{Gutsev00,Gonzales00,Kudin07}


The $GW$ approximation is unique, but due to its high computational costs, there are various $GW$ schemes and variants. Generally, there are two approaches. One approach is to vary the self-consistency level in the $GW$ approximation. The $GW$ self-consistent levels from the lowest to the highest include the perturbative non-self-consistent (one-shot) $GW$ ($G_{0}W_{0}$) scheme, the eigenvalue self-consistent $GW$ (ev$GW$) scheme (with two types $G_{n}W_{0}$ and $G_{n}W_{n}$, which update eigenvalues only in $G$ and in both $G$ and $W$, respectively), the QP self-consistent $GW$ (QS$GW$) scheme using a static and Hermitian approximation to the $GW$ self-energy, and the fully self-consistent $GW$ (SC$GW$) scheme.~\cite{Bruneval14} Generally, as the $GW$ self-consistency level increases, the $GW$ approximation depends less on the mean-field starting point and becomes more conserving with respect to particle number, momentum, and energy. However, the higher $GW$ self-consistency level does not necessarily give more accurate QP energies because vertex corrections are missing in the $GW$ approximation. For example, SC$GW$ and QS$GW$ systematically overestimate the bandgaps of solids,~\cite{Grumet18} displaying worse performance than ev$GW$, which currently provides the best balance between accuracy and efficiency for solids.~\cite{Shishkin07}


The other approach is to vary the amount of EXX in the $GW$ starting point to reduce the self-interaction error by (semi-)local xc functionals. Typically, the $G_{0}W_{0}$ scheme chooses this approach to obtain good results at low computational costs. However, the predictive power of this approach is questionable, since the optimal amount of EXX in the $GW$ starting point is strongly system-dependent. For example, for extended systems, the reported values for the optimal amount of EXX are narrowly spread between 0\% and 25\%,~\cite{Fuchs07}, while for finite systems, they are widely spread between 25\% and 100\%.~\cite{Korbel14,Bruneval13,Kaplan16,Rostgaard10}


The purpose of this work is to evaluate the performance of the $GW$ approximation on the electronic structure of small oxide molecules containing 3$d$ transition metals. To this end, we calculate the QP spectra of closed- and open-shell molecular anions with partially and completely filled 3$d$ shells, ScO$^{-}$, TiO$^{-}$, CuO$^{-}$, and ZnO$^{-}$, using various levels of $GW$ theory. There are a few reasons why we chose these molecular systems: (i) their anion PES data is available,~\cite{Wu98,Wu97a,Wu97b,Moravec01} (ii) CuO$^{-}$ and ZnO$^{-}$ are molecular analogs to bulk Cu$_{2}$O and ZnO, respectively, which are challenging systems for the $GW$ method,~\cite{Bruneval06,Isseroff12} and (iii) shallow and deep 3$d$ states are measured in TiO$^{-}$ and CuO$^{-}$, respectively.

This article is organized as follows: First, we give a brief introduction to the $GW$ approximation and its implementation in the framework of quantum chemistry. Second, we present various convergence test results and show that care should be taken to obtain reliable and reproducible QP energies of finite systems from Gaussian-based $GW$ implementations. Third, we assess various $GW$ schemes, focusing on ionization energies and 3$d$-electron binding energies, and conclude that the $G_{n}W_{0}@\text{PBE}$ scheme gives the best performance among $GW$ schemes considered in this work in terms of accuracy and efficiency. Last, we discuss the origin of seemingly conflicting $GW$ results for finite systems in the literature.

\section{Theoretical Background}

In this section, we briefly review the $GW$ approximation and its implementation using local-orbital basis sets. This section contains only a minimal number of equations, which will be needed later. More details can be found in Refs.~\onlinecite{Bruneval16,vanSetten15,vanSetten13,Tiago06}. Generally, we follow the notation in the MOLGW implementation paper~\cite{Bruneval16} for consistency: (i) Hartree atomic units are used in all equations, (ii) The complex conjugate notation is not used for wavefunctions, because they are real in finite systems, (iii) State indices $i$ and $j$ run over only occupied states, $a$ and $b$ run over only empty (virtual) states, and $m$ and $n$ run over all states, (iv) The response function is refered to as the polarizability instead of the susceptibility, and (v) $\chi$ is used for the polarizability instead of $P$ and $\Pi$.


\subsection{$GW$ Approximation}

In Hedin's $GW$ approximation, the non-local, dynamical, and non-Hermitian self-energy $\Sigma^{\sigma}$ at frequency $\omega$ is given by
\begin{equation}
\Sigma^{\sigma}(\mathbf{r},\mathbf{r'},\omega) = \frac{i}{2\pi} \int d\omega' e^{i\eta\omega'} G^{\sigma}(\mathbf{r},\mathbf{r'},\omega+\omega') W(\mathbf{r'},\mathbf{r},\omega'), \label{eq:Sigma}
\end{equation}
where $\sigma$ is the spin channel ($\uparrow$ or $\downarrow$), $G^{\sigma}$ is the time-ordered one-particle Green's function, $W$ is the dynamically screened Coulomb interaction, and $\eta$ is a positive infinitesimal.


The self-energy in Eq.~(\ref{eq:Sigma}) can be calculated from first principles by solving the coupled Hedin's equations in order. One starts by constructing the one-particle Green's function using the one-electron eigenvalues $\epsilon^{\sigma}_{m}$ and corresponding wavefunctions $\varphi^{\sigma}_{m}(\mathbf{r})$ obtained from the Hartree or mean-field approximation: 
\begin{equation}
G^{\sigma}(\mathbf{r},\mathbf{r'},\omega) = \sum_{i} \frac{\varphi^{\sigma}_{i}(\mathbf{r})\varphi^{\sigma}_{i}(\mathbf{r'})}{\omega - \epsilon^{\sigma}_{i} - i\eta} + \sum_{a} \frac{\varphi^{\sigma}_{a}(\mathbf{r})\varphi^{\sigma}_{a}(\mathbf{r'})}{\omega - \epsilon^{\sigma}_{a} + i\eta}, \label{eq:G}
\end{equation}
where $i$ runs over occupied states and $a$ runs over empty states. Note that $G^{\sigma}$ in Eq.~(\ref{eq:G}) is not the interacting (dressed) Green's function, but the non-interacting (bare) one, which are conventionally denoted by $G^{\sigma}$ and $G^{\sigma}_{0}$, respectively. In this work, we use the subscript 0 to distinguish the non-self-consistent $GW$ scheme from the self-consistent one.


Using the one-particle Green's function in Eq.~(\ref{eq:G}), one can successively obtain the non-interacting (irreducible) polarizability $\chi_{0}$ and the interacting (reducible) polarizability $\chi = \chi_{0} [1 - v \chi_{0}]^{-1}$ within the RPA, the screened Coulomb interaction, and the self-energy:
\begin{equation}
\chi_{0} = -i \sum_{\sigma} G^{\sigma} G^{\sigma}, \label{eq:chi0}
\end{equation}
\begin{equation}
W = v + v \chi_{0} W = v + v \chi_{0} v + v \chi_{0} v \chi_{0} v + ... = v + v \chi v, \label{eq:W}
\end{equation}
\begin{equation}
\Sigma^{\sigma} = i G^{\sigma} W = i G^{\sigma} (v + v \chi v) = \Sigma^{\sigma}_{\text{x}} + \Sigma^{\sigma}_{\text{c}}, \label{eq:Sigma.short}
\end{equation}
where $v$ denotes the bare (unscreened) Coulomb interaction $v(\mathbf{r},\mathbf{r'}) = 1 / |\mathbf{r} - \mathbf{r'}|$, $\Sigma_{\text{x}}$ is the exchange part of the self-energy, and $\Sigma_{\text{c}}$ is the correlation part of the self-energy. Note that in Eqs.~(\ref{eq:chi0}), (\ref{eq:W}), and (\ref{eq:Sigma.short}), space and frequency variables ($\mathbf{r}$, $\mathbf{r'}$, $\omega$) are omitted for simplicity, and $\chi_{0}(\omega)$, $\chi(\omega)$, $\Sigma^{\sigma}(\omega)$, and $\Sigma^{\sigma}_{\text{c}}(\omega)$ are dynamic, whereas $v$ and $\Sigma^{\sigma}_{\text{x}}$ are static. Note also that $W$ is obtained without using the dielectric matrix.


Using the real part (Re) of the self-energy in Eq.~(\ref{eq:Sigma.short}) and the first-order perturbation theory, one can obtain the (diagonal) QP equation: 
\begin{equation}
\epsilon^{\text{G}_{0}\text{W}_{0},\sigma}_{m} = \epsilon^{\sigma}_{m} + \langle \varphi^{\sigma}_{m} | \text{Re}\Sigma^{\sigma}(\epsilon^{\text{G}_{0}\text{W}_{0},\sigma}_{m}) - v^{\sigma}_{\text{xc}} | \varphi^{\sigma}_{m} \rangle, \label{eq:G0W0.QP.eq.non.lin}
\end{equation}
where $\epsilon^{\text{G}_{0}\text{W}_{0},\sigma}_{m}$ are the perturbative one-shot $GW$ QP energies and $v^{\sigma}_{\text{xc}}$ is the xc potential. Experimentally, $\epsilon^{\text{G}_{0}\text{W}_{0},\sigma}_{m}$ correspond to vertical IEs and EAs in PES and IPES, respectively. Theoretically, $\epsilon^{\text{G}_{0}\text{W}_{0},\sigma}_{m}$ correspond to the positions of poles of the Green's function in the spectral (Lehmann) representation and thereby to the positions of QP peaks and plasmon satellites in the corresponding spectral function $A^{\sigma}$:
\begin{equation}
A^{\sigma}_{mm}(\mathbf{r},\mathbf{r'},\omega)= \frac{1}{\pi} |\text{Im}G^{\sigma}_{mm}(\mathbf{r},\mathbf{r'},\omega)|, \label{eq:A}
\end{equation}
where $A^{\sigma}_{mm}$ are the diagonal elements of the spectral function, $G^{\sigma}_{mm}$ are the diagonal elements of the Green's function, and Im represents the imaginary part. Note that $A^{\sigma}_{mm}$ give the local density of states.


$G^{\sigma}$ in Eq.~(\ref{eq:A}) is the interacting Green's function, whereas $G^{\sigma}$ in Eq.~(\ref{eq:G}) is the non-interacting one. In other words, by plugging $G^{\sigma}$ in Eq.~(\ref{eq:G}) into Eq.~(\ref{eq:A}) after replacing $\epsilon^{\sigma}_{m}$, where $m$ = $i$ or $a$, by $\epsilon^{\sigma}_{m} + \langle \varphi^{\sigma}_{m} | \Sigma^{\sigma}(\omega) - v^{\sigma}_{\text{xc}} | \varphi^{\sigma}_{m} \rangle$, one can find that $A^{\sigma}_{mm}$ have Lorentzian peaks at
\begin{equation}
\omega = \epsilon^{\sigma}_{m} + \langle \varphi^{\sigma}_{m} | \text{Re}\Sigma^{\sigma}(\omega) - v^{\sigma}_{\text{xc}} | \varphi^{\sigma}_{m} \rangle, \label{eq:A.peak}
\end{equation}
which shows that solving the QP equation in Eq.~(\ref{eq:G0W0.QP.eq.non.lin}) and locating the peak positions in the spectral function in Eq.~(\ref{eq:A}) are equivalent ways of obtaining the QP energies. 

The QP equation in Eq.~(\ref{eq:G0W0.QP.eq.non.lin}) is non-linear, because $\Sigma^{\sigma}$ depends on $\epsilon^{\text{G}_{0}\text{W}_{0},\sigma}_{m}$, so it should be solved numerically. Additionally, Hedin equations are coupled, because $W$ and $\Sigma^{\sigma}$ depend on $G^{\sigma}$, so they should be solved self-consistently. Multiple ways to numerically solve the non-linear QP equation and to iteratively solve the coupled Hedin equations will be discussed later.

\subsection{Self-Consistent Field Method} \label{sec:SCF.Method}

In order to obtain the ingredients for the one-particle Green's function in Eq.~(\ref{eq:G}) using local-orbital basis sets, molecular orbitals (MOs) and corresponing MO energies are used as one-electron wavefunctions and corresponding eigenvalues. MOs are expanded as a linear combination of atomic orbitals (AOs) $\phi_{\mu}$:
\begin{equation}
\varphi^{\sigma}_{m}(\mathbf{r}) = \sum_{\mu} C^{\sigma}_{\mu m} \phi_{\mu}(\mathbf{r}), \label{eq:MO}
\end{equation}
where $C^{\sigma}_{\mu m}$ are MO expansion coefficients. In MOLGW, atom-centered (contracted) Gaussian orbitals are used as AOs.

The MO coefficients in Eq.~(\ref{eq:MO}) and MO energies can be obtained by solving the generalized Kohn-Sham (gKS) equation [the Hartree-Fock--Kohn-Sham scheme~\cite{Seidl96} for (semi-)local functionals, hybrid functionals, and HF]:
\begin{equation}
\mathbf{H}^{\sigma} \mathbf{C}^{\sigma} = \mathbf{S} \mathbf{C}^{\sigma} \mathbf{\epsilon}^{\sigma}, \label{eq:gKS}
\end{equation}
where $\mathbf{C}^{\sigma}$ is a matrix of MO coefficients, $\mathbf{\epsilon}^{\sigma}$ is a diagonal matrix of MO energies, $\mathbf{S}$ is the AO overlap matrix with elements:
\begin{equation}
S_{\mu\nu} = \int d\mathbf{r} \phi_{\mu}(\mathbf{r}) \phi_{\nu}(\mathbf{r}), \label{eq:S}
\end{equation}
and $\mathbf{H}^{\sigma}$ is the Hamiltonian matrix with elements:
\begin{equation}
H^{\sigma}_{\mu\nu} = T_{\mu\nu} + V_{\text{ext},\mu\nu} + J_{\mu\nu} - \alpha K^{\sigma}_{\mu\nu} + (1 - \alpha) V^{\text{PBE},\sigma}_{\text{x},\mu\nu} + V^{\text{PBE},\sigma}_{\text{c},\mu\nu}, \label{eq:H}
\end{equation}
where $T$, $V_{\text{ext}}$, $J$, and $K^{\sigma}$ are the kinetic energy, external potential energy, Hartree, and Fock exchange terms, respectively, $V^{\sigma}_{\text{x}}$ and $V^{\sigma}_{\text{c}}$ are the exchange and correlation potentials, respectively, and $\alpha$ is the fraction of EXX in hybrid functionals that will be introduced later. 


We briefly explain only a few terms in the Hamiltonian matrix in Eq.~(\ref{eq:H}), which will be needed later. The matrix elements of the Hartree term in Eq.~(\ref{eq:H}) are given by
\begin{equation}
J_{\mu\nu} = \sum_{\lambda\tau} (\mu\nu|\lambda\tau) \sum_{\sigma} D^{\sigma}_{\lambda\tau}, \label{eq:J}
\end{equation}
where $(\mu\nu|\lambda\tau)$ are the 4-center two-electron Coulomb repulsion integrals:
\begin{equation}
(\mu\nu|\lambda\tau) = \iint d\mathbf{r} d\mathbf{r'} \phi_{\mu}(\mathbf{r}) \phi_{\nu}(\mathbf{r}) \frac{1}{|\mathbf{r} - \mathbf{r'}|} \phi_{\lambda}(\mathbf{r'}) \phi_{\tau}(\mathbf{r'}), \label{eq:AO.4.ERI}
\end{equation}
and $\mathbf{D}^{\sigma}$ is the density matrix with elements:
\begin{equation}
D^{\sigma}_{\mu\nu} = \sum_{m} f^{\sigma}_{m} C^{\sigma}_{\mu m} C^{\sigma}_{\nu m}, \label{eq:D}
\end{equation}
where $f^{\sigma}$ is the occupation number (0 or 1). The matrix elements of the Fock exchange term in Eq.~(\ref{eq:H}) are given by
\begin{equation}
K^{\sigma}_{\mu\nu} = \sum_{\lambda\tau} D^{\sigma}_{\lambda\tau} (\mu\lambda|\tau\nu). \label{eq:K}
\end{equation}
The exchange and correlation potentials in Eq.~(\ref{eq:H}) depend on the density $\rho^{\sigma}$ (and the density gradient $\nabla\rho^{\sigma}$):
\begin{equation}
\rho^{\sigma}(\mathbf{r}) = \sum_{\mu\nu} D^{\sigma}_{\mu\nu} \phi_{\mu}(\mathbf{r}) \phi_{\nu}(\mathbf{r}). \label{eq:rho}
\end{equation}

The gKS equation in Eq.~(\ref{eq:gKS}) (the restricted Roothaan-Hall or unrestricted Pople-Nesbet equations) should be solved using the SCF method, because $J$, $K^{\sigma}$,  $V^{\sigma}_{\text{x}}$, and $V^{\sigma}_{\text{c}}$ in Eq.~(\ref{eq:H}) depend on the density matrix in Eq.~(\ref{eq:D}), as shown in Eqs.~(\ref{eq:J}), (\ref{eq:K}), and (\ref{eq:rho}).

\subsection{$GW$ Self-Energy} \label{sec:GW.Self.Energy}

In order to obtain the ingredients for the interacting polarizability in Eq.~(\ref{eq:W}), one should solve the Casida equation in matrix form:
\begin{equation}
 \begin{pmatrix*}[r]
  \mathbf{A} & \mathbf{B} \\
  -\mathbf{B} & -\mathbf{A}
 \end{pmatrix*}
 \begin{pmatrix}
  X^{s} \\
  Y^{s}
 \end{pmatrix}
 =
 \Omega_{s}
 \begin{pmatrix}
  X^{s} \\
  Y^{s}
 \end{pmatrix}, \label{eq:Casida}
\end{equation}
where $\mathbf{A}$ and $\mathbf{B}$ are the resonant and coupling matrices, respectively, and $\Omega_{s}$ and $(X^{s}, Y^{s})$ are the eivenvalues (the neutral two-particle excitation energies) and corresponding eigenvectors, respectively. The matrix elements in $\mathbf{A}$ and $\mathbf{B}$ are given by
\begin{align}
A^{jb\sigma'}_{ia\sigma} &= (\epsilon^{\sigma}_{a} -\epsilon^{\sigma}_{i}) \delta_{ij} \delta_{ab} \delta_{\sigma\sigma'} + (ia\sigma|jb\sigma') + f^{jb\sigma'}_{\text{xc},ia\sigma}, \label{eq:Casida.A} \\
B^{jb\sigma'}_{ia\sigma} &= (ia\sigma|bj\sigma') + f^{bj\sigma'}_{\text{xc},ia\sigma}, \label{eq:Casida.B}
\end{align}
where $i$ and $j$ are for occupied states, $a$ and $b$ are for empty states, $f_{\text{xc}}$ is the time-dependent density-functional theory (TDDFT) xc kernel, and $(ia\sigma|jb\sigma')$ are the 4-orbital two-electron Coulomb repulsion integrals: 
\begin{equation}
(ia\sigma|jb\sigma') = \iint d\mathbf{r} d\mathbf{r'} \varphi^{\sigma}_{i}(\mathbf{r}) \varphi^{\sigma}_{a}(\mathbf{r}) \frac{1}{|\mathbf{r} - \mathbf{r'}|} \varphi^{\sigma'}_{j}(\mathbf{r'}) \varphi^{\sigma'}_{b}(\mathbf{r'}). \label{eq:MO.4.ERI}
\end{equation}
In this work, we used the RPA by setting $f_{\text{xc}} = 0$. Note that Ref.~\onlinecite{vanSetten13} showed that TDDFT and RPA polarizabilities make a difference of $\sim$0.1~eV in $G_{0}W_{0}@\text{PBE}$ QP HOMO energy. Note also that the dimension of the Casida matrix in Eq.~(\ref{eq:Casida}) scales as $O(N^{2})$ with $N$ being the system size, so building and completely diagonalizing the Casida matrix scale as $O(N^{4})$ and $O(N^{6})$, respectively. The MO integrals in Eq.~(\ref{eq:MO.4.ERI}) are transformed from the AO integrals in Eq.~(\ref{eq:AO.4.ERI}) through the AO-MO integral transformation:
\begin{equation}
(ia\sigma|jb\sigma') = \sum_{\mu\nu\lambda\tau} C^{\sigma}_{\mu i} C^{\sigma}_{\nu a} C^{\sigma'}_{\lambda j} C^{\sigma'}_{\tau b} (\mu\nu|\lambda\tau), \label{eq:AO.to.MO}
\end{equation}
which scales as $O(N^{5})$. Note that this integral transformation is a bottleneck in Gaussian-based $GW$ and MP2 calculations.


Diagonalizing the Casida matrix in Eq.~(\ref{eq:Casida}) yields eigenvalues $\Omega_{s}$ and eigenvectors $(X^{s}, Y^{s})$. In MOLGW, the diagonalization is performed without using the Tamm-Dancoff approximation (TDA), which sets $\mathbf{B}$ to zero, but efficiently using the so-called beyond-TDA method.~\cite{Sander2015,Bruneval16,Byun2017}. Using $\Omega_{s}$ and $(X^{s}, Y^{s})$, one can construct the spectral representation of the interacting polarizability $\chi({\omega})$.~\cite{Tiago06,vanSetten15,vanSetten13} From $\chi({\omega})$ and Eq.~(\ref{eq:W}), one can obtain the spectral representation of the screened Coulomb interaction $W({\omega})$.~\cite{Tiago06,vanSetten13,vanSetten15,Bruneval16} Using $W({\omega})$ and Eq.~(\ref{eq:Sigma.short}), and analytically performing the convolution of $G^{\sigma}(\omega)$ and $W(\omega)$ in the frequency domain, one can obtain the exchange and correlation parts of the $GW$ self-energy $\Sigma^{\sigma}_{\text{x}}$ and $\Sigma^{\sigma}_{\text{c}}(\omega)$, respectively, whose diagonal matrix elements are given by
\begin{align}
\langle \varphi^{\sigma}_{m} | \Sigma^{\sigma}_{\text{x}} | \varphi^{\sigma}_{m} \rangle & = - \sum_{i} (mi\sigma|im\sigma), \label{eq:Sigma.x} \\
\langle \varphi^{\sigma}_{m} | \Sigma^{\sigma}_{\text{c}} (\omega) | \varphi^{\sigma}_{m} \rangle &= \sum_{is} \frac{w^{s}_{mi\sigma} w^{s}_{mi\sigma}}{\omega - \epsilon^{\sigma}_{i} + \Omega_{s} - i\eta}, \nonumber \\
&+ \sum_{as} \frac{w^{s}_{ma\sigma} w^{s}_{ma\sigma}}{\omega - \epsilon^{\sigma}_{a} - \Omega_{s} + i\eta}, \label{eq:Sigma.c}
\end{align}
where $i$ runs over occupied states, $a$ runs over empty states, $s$ runs over all excitations, and $w^{s}_{mn\sigma}$ are given by
\begin{equation}
w^{s}_{mn\sigma} = \sum_{ia\sigma'} (mn\sigma|ia\sigma') (X^{s}_{ia\sigma'} + Y^{s}_{ia\sigma'}). \label{eq:residue}
\end{equation}


Note that unlike the plasmon-pole approximation (PPA), the analytic continuation method, and the contour deformation technique,~\cite{vanSetten15,Golze18} the fully analytic method employed in RGWBS,~\cite{Tiago06} TURBOMOLE, and MOLGW gives the exact $GW$ self-energy at all frequency points because it does not rely on any approximation and numerical parameter.

\subsection{Spin Contamination}

In unresticted HF and KS calculations for open-shell systems, the expectation value of the total angular momentum $\langle S^{2} \rangle$ is given by 
\begin{equation}
\langle S^{2} \rangle = S(S+1) + N_{\downarrow} - \sum_{i}^{N_{\uparrow}} \sum_{j}^{N_{\downarrow}} |\langle \varphi_{i}^{\uparrow} | \varphi_{j}^{\downarrow} \rangle|^{2}, \label{eq:S2}
\end{equation}
where $N_{\uparrow}$ and $N_{\downarrow}$ are the numbers of $\uparrow$- and $\downarrow$-spin electrons, respectively, and $S$ is (${N_{\uparrow} - N_{\downarrow}}$)/2 with $N_{\uparrow} > N_{\downarrow}$. The last two terms on the right side of Eq.~(\ref{eq:S2}) are called the spin contamination, which is non-negative.~\cite{Cohen07,Menon08} The spin contamination becomes large when a ground state is mixed with (contaminated by) excited states.


In restricted calculations for closed-shell systems, the SCF cycle always converges to a global minimum and the spin contamination is zero for all (semi-)local and hybrid functionals as well as HF. In unrestricted calculations for open-shell systems, the SCF convergence and the spin contamination depend on EXX amount and basis size. For (semi-)local functionals, the SCF cycle almost always converges to a global minimum and the spin contamination is small [generally smaller than $\sim$10\% of $S(S+1)$]. For hybrid functionals and HF, there is a chance (which increases with EXX amount and basis size) that the SCF cycle fails, does not converge, or converges to local minima or the spin contamination is large. 

There are a few points to note about the spin contamination. First, the spin contamination is just an indicator for the SCF convergence error, therefore, a small spin contamination does not guarantee the correct SCF convergence. Second, that the spin contamination generally raises, but sometimes lowers the gKS total energy, so the lowest gKS total energy does not guarantee the correct SCF convergence, either. Last, the spin contamination and the SCF cycle are independent of each other. For example, the SCF cycle can converge quickly with large spin contamination or slowly with small spin contamination. 



\subsection{Auxiliary Basis Sets and Multi-thread Parallelization}

In Gaussian-based $GW$, the 4-center integrals $(\mu\nu|\lambda\tau)$ in Eq.~(\ref{eq:AO.4.ERI}), which scale as $O(N^{4})$, are a common bottleneck in gKS and $GW$ parts in terms of compute time and memory storage. One way to reduce the bottleneck is the resolution-of-identity (RI) approximation (the density-fitting approximation), which expands the product of basis functions $\phi_{\mu}(\mathbf{r}) \phi_{\nu}(\mathbf{r})$ as a linear combination of auxiliary basis functions $\phi_{\text{P}}(\mathbf{r})$.~\cite{Ren12,Hill08,Hill12} There are two types of the RI approximation: RI-V using a Coulomb metric and RI-SVS using an overlap metric. For example, FIESTA uses both RI-V and RI-SVS, whereas MOLGW uses only RI-V, which is known to be superior to RI-SVS. Within RI-V, the 4-center integrals $(\mu\nu|\lambda\tau)$ in Eq.~(\ref{eq:AO.4.ERI}) approximate to
\begin{equation}
(\mu\nu|\lambda\tau) \approx \sum_{PQ} (\mu\nu|P) (P|Q)^{-1} (Q|\lambda\tau), \label{eq:RI}
\end{equation}
where $P$ and $Q$ run over auxiliary basis functions, $(\mu\nu|P)$ and $(Q|\lambda\tau)$ are the 3-center integrals, and $(P|Q)$ are the 2-center integrals.


RI can be applied to both gKS [$J$ and $K^{\sigma}$ in Eqs.~(\ref{eq:J}) and (\ref{eq:K})] and $GW$ [$\mathbf{A}$, $\mathbf{B}$, $\Sigma^{\sigma}_{\text{x}}$, and $\Sigma^{\sigma}_{\text{c}}(\omega)$ in Eqs.~(\ref{eq:Casida.A}), (\ref{eq:Casida.B}), (\ref{eq:AO.to.MO}), (\ref{eq:Sigma.x}), (\ref{eq:Sigma.c}), and (\ref{eq:residue})] parts. In this work, we refer to RI applied to one (both) of them as a half (full) RI method. For example, FIESTA uses a half RI method, whereas MOLGW uses a full RI method. In this work, we observed that a full RI method in MOLGW reduces both compute time and memory storage by about the number of basis functions (by $\sim$100 times as shown in Table~\ref{tab:Occupied.Empty.States}).


RI is an approximation, so it causes an error. There are mixed results for the RI error in the literature, ranging from $\sim$1 meV to $\sim$0.1 eV, because different molecular systems, molecular orbitals, levels of theory (DFT vs $GW$), xc functionals (PBE vs HF), and basis sets are used to evaluate the quality of RI.~\cite{Korbel14,vanSetten15}



Another way to reduce the bottleneck without causing an error is the parallelization. We parallelized the 4-center integrals in Eqs.~(\ref{eq:J}), (\ref{eq:K}), (\ref{eq:Casida.A}), (\ref{eq:Casida.B}), (\ref{eq:AO.to.MO}), (\ref{eq:Sigma.x}), (\ref{eq:Sigma.c}), and (\ref{eq:residue}) [as well as other bottlenecks, such as the integral transformation in Eq.~(\ref{eq:AO.to.MO}) and the correlation part of the self-energy in Eq.~(\ref{eq:Sigma.c})] using Open Multi-Processing (OpenMP), which consumes much less memory than Message Passing Interface by using shared-memory threads. The performance gain by our OpenMP parallelization is shown in supplementary material. We also optimized our OpenMP implementation to reduce Non-Uniform Memory Access (NUMA) effects in modern multi-core processors by enhancing the memory bandwidth and reducing the memory latency. Our OpenMP implementation in MOLGW 1.F has recently been merged into MOLGW 2.A. 



\subsection{$G_{0}W_{0}$ Quasiparticle Energy} \label{sec:background.G0W0.QP.energy}

In this work, we used three methods to obtain $\epsilon^{\text{G}_{0}\text{W}_{0}}_{m}$, as it is practically impossible to obtain unique, correct, and accurate $\epsilon^{\text{G}_{0}\text{W}_{0}}_{m}$ for all energy levels of all molecular systems using a single method, which will be discussed in detail later. Note that in the following, the spin channel $\sigma$ and the real part Re are omitted for simplicity.


The first method is to linearize the non-linear QP equation in Eq.~(\ref{eq:G0W0.QP.eq.non.lin}):
\begin{equation}
\epsilon^{\text{G}_{0}\text{W}_{0}}_{m} \approx \epsilon_{m} + Z^{\text{linear}}_{m} \langle \varphi_{m} | \Sigma(\epsilon_{m}) - v_{\text{xc}} | \varphi_{m} \rangle \equiv \epsilon^{\text{G}_{0}\text{W}_{0},\text{linear}}_{m}, \label{eq:G0W0.QP.eq.lin}
\end{equation}
where $\epsilon^{\text{G}_{0}\text{W}_{0},\text{linear}}_{m}$ is the perturbative one-shot QP energy obtained from the linearization, and $Z^{\text{linear}}_{m}$ is the QP renormalization factor for the linearization:
\begin{equation}
Z^{\text{linear}}_{m}(\epsilon_{m}) = \frac{1}{1 - \frac{\partial}{\partial \omega} \langle \varphi_{m} | \Sigma(\omega) | \varphi_{m} \rangle |_{\omega = \epsilon_{m}}}, \label{eq:Z.lin}
\end{equation}
where the derivative of the self-energy is obtained from the finite difference method using two frequency points at $\epsilon_{m} \pm \Delta\omega$ with $\Delta\omega$ being the frequency grid spacing, which is set to 0.001~Ha in this work.

There are a few points to note about the linearization method. First, one can choose different frequency points for the finite difference method (e.g. $\epsilon_{m} \pm 0.1$~eV and $\epsilon_{m} \pm 0.5$~eV in Refs.~\onlinecite{Maggio17,Hybertsen86}, respectively). In the PPA $G_{0}W_{0}$ method, different frequency points give similar results for $\epsilon^{\text{G}_{0}\text{W}_{0},\text{linear}}_{m}$, because PPA makes $\langle \varphi_{m} | \Sigma_{\text{c}} (\omega) | \varphi_{m} \rangle$ in Eq.~(\ref{eq:Sigma.c}) smooth around $\epsilon_{m}$ by drastically reducing the number of self-energy poles.~\cite{Hybertsen86} However, in the full-frequency $G_{0}W_{0}$ method, different frequency points can give very different results for $\epsilon^{\text{G}_{0}\text{W}_{0},\text{linear}}_{m}$ when the finite difference method fails due to weak self-energy poles around $\epsilon_{m}$, which will be discussed later. Second, the derivative of the self-energy can be evaluated analytically,~\cite{vanSetten13} but it is not used in this work. Last, $Z^{\text{linear}}_{m}$ in Eq.~(\ref{eq:Z.lin}) is slightly different from that in Ref.~\onlinecite{vanSetten15},
\begin{equation}
Z_{m}(\epsilon^{\text{G}_{0}\text{W}_{0}}_{m}) = \frac{1}{1 - \frac{\partial}{\partial \omega} \langle \varphi_{m} | \Sigma(\omega) | \varphi_{m} \rangle |_{\omega = \epsilon^{\text{G}_{0}\text{W}_{0}}_{m}}}. \label{eq:Z.GW100}
\end{equation}
The derivative of the self-energy is evalutated at $\omega = \epsilon_{m}$ and $\omega = \epsilon^{\text{G}_{0}\text{W}_{0}}_{m}$ in $Z^{\text{linear}}_{m}$ and $Z_{m}$, respectively. Generally, $Z_{m}$ is smaller than $Z^{\text{linear}}_{m}$ because the self-energy has a steeper slope at $\omega = \epsilon^{\text{G}_{0}\text{W}_{0}}_{m}$ than at $\omega = \epsilon_{m}$. $Z_{m}$ represents the QP weight (the pole residue of the Green's function), which equals the area under the Lorentzian QP peak and depends on the spectral weight transfer from the QP peak to plasmon satellites and the incoherent background.


The second method is to graphically solve the non-linear QP equation in Eq.~(\ref{eq:G0W0.QP.eq.non.lin}) using the secant (quasi-Newton) method. In this work, we refer to $\epsilon^{\text{G}_{0}\text{W}_{0}}_{m}$ obtained from the graphical solution as $\epsilon^{\text{G}_{0}\text{W}_{0},\text{graph}}_{m}$. Note that the above linearization method corresponds to the first step of the secant method. Note also that when the non-linear QP equation in Eq.~(\ref{eq:G0W0.QP.eq.non.lin}) has multiple solutions, the secant method gives only one of them, which depends strongly on the choice of $\eta$.


The last method is to use the position of the QP peak with the highest spectral weight in the spectral function $A_{mm}$ in Eq.~(\ref{eq:A}). In this work, we refer to $\epsilon^{\text{G}_{0}\text{W}_{0}}_{m}$ obtained from the spectral function as $\epsilon^{\text{G}_{0}\text{W}_{0},\text{spect}}_{m}$ and define $Z^{\text{spect}}_{m}$ by replacing $Z^{\text{linear}}_{m}$ and $\epsilon^{\text{G}_{0}\text{W}_{0},\text{linear}}_{m}$ in Eq.~(\ref{eq:G0W0.QP.eq.lin}) by $Z^{\text{spect}}_{m}$ and $\epsilon^{\text{G}_{0}\text{W}_{0},\text{spect}}_{m}$, respectively. Note that we searched for the QP peak at $0 < Z^{\text{spect}}_{m} < 1$ using the peak height instead of the spectral weight (the area under the peak) due to the practical difficulty of determining the peak range. Note also that the highest spectral weight gives the largest $Z_{m}$ in Eq.~(\ref{eq:Z.GW100}) because $Z_{m}$ represents the spectral weight, as explained above.

\subsection{$G_{n}W_{0}$ and $G_{n}W_{n}$ Quasiparticle Energy} \label{sec:background.evGW.QP.energy}

As introduced in Section~\ref{sec:introduction}, there are various levels of self-consistency in the $GW$ approximation (from the lowest to the highest): $G_{0}W_{0}$, $G_{n}W_{0}$, $G_{n}W_{n}$, QS$GW$, and SC$GW$. In this work, we used $G_{n}W_{0}$ and $G_{n}W_{n}$ for simplicity, efficiency, and stability. $G_{n}W_{0}$ updates only gKS eigenvalues in $G^{\sigma}$ [$\epsilon^{\sigma}_{i}$ and $\epsilon^{\sigma}_{a}$ in Eq.~(\ref{eq:Sigma.c})], whereas $G_{n}W_{n}$ updates gKS eigenvalues in $G^{\sigma}$ and $\mathbf{A}$ [$\epsilon^{\sigma}_{i}$ and $\epsilon^{\sigma}_{a}$ in Eq.~(\ref{eq:Casida.A})] as well as Casida eigenvalues in $W$ [$\Omega_{s}$ in Eq.~(\ref{eq:Sigma.c})]. Therefore, $G_{n}W_{n}$ is computationally more expensive than $G_{n}W_{0}$ by the time to build and completely diagonalize the RPA Casida matrix in Eq.~(\ref{eq:Casida}). Note that $G_{n}W_{n}$ can be viewed as a diagonal approximation to QS$GW$.


In this work, we obtained $G_{n}W_{0}$ and $G_{n}W_{n}$ QP energies ($\epsilon^{\text{G}_{n}\text{W}_{0}}_{m}$ and $\epsilon^{\text{G}_{n}\text{W}_{n}}_{m}$, respectively) by iterating the recurrence relations ($n \ge 3$):
\begin{align}
\epsilon^{\text{evGW},1}_{m} &= \epsilon_{m} + Z^{\text{evGW}} \langle \varphi_{m} | \Sigma(\epsilon_{m}) - v_{\text{xc}} | \varphi_{m} \rangle, \label{eq:evGW.QP.eq.1} \\
\epsilon^{\text{evGW,2}}_{m} &= \epsilon^{\text{evGW,1}}_{m} \nonumber \\
&+ Z^{\text{evGW}} \langle \varphi_{m} | \Sigma(\epsilon^{\text{evGW,1}}_{m}) - \Sigma(\epsilon_{m}) | \varphi_{m} \rangle, \label{eq:evGW.QP.eq.2} \\
\epsilon^{\text{evGW},n}_{m} &= \epsilon^{\text{evGW},n-1}_{m} \nonumber \\
&+ Z^{\text{evGW}} \langle \varphi_{m} | \Sigma(\epsilon^{\text{evGW},n-1}_{m}) - \Sigma(\epsilon^{\text{evGW},n-2}_{m}) | \varphi_{m} \rangle, \label{eq:evGW.QP.eq.n}
\end{align}
where $\epsilon^{\text{evGW},n}_{m}$ is $\epsilon^{\text{G}_{n}\text{W}_{0}}_{m}$ or $\epsilon^{\text{G}_{n}\text{W}_{n}}_{m}$, and $Z^{\text{evGW}} = 1$. Summing up Eqs.~(\ref{eq:evGW.QP.eq.1}), (\ref{eq:evGW.QP.eq.2}), and (\ref{eq:evGW.QP.eq.n}), we get
\begin{equation}
\epsilon^{\text{evGW},n}_{m} = \epsilon_{m} + \langle \varphi_{m} | \Sigma(\epsilon^{\text{evGW},n-1}_{m}) - v_{\text{xc}} | \varphi_{m} \rangle, \label{eq:evGW.QP.eq}
\end{equation}
which we refer to as the ev$GW$ QP equation in this work. When the ev$GW$ convergence is reached ($\epsilon^{\text{evGW},n}_{m} = \epsilon^{\text{evGW},n-1}_{m} =  \epsilon^{\text{evGW},\infty}_{m}$, where $\epsilon^{\text{evGW},\infty}_{m}$ are converged ev$GW$ QP energies), the ev$GW$ QP equation in Eq.~(\ref{eq:evGW.QP.eq}) becomes similar to the $G_{0}W_{0}$ QP equation in Eq.~(\ref{eq:G0W0.QP.eq.non.lin}). 


Whereas most $GW$ codes use $0 < Z^{\text{evGW}} < 1$,~\cite{Shishkin07,Blase11,Veril18} MOLGW uses $Z^{\text{evGW}} = 1$. Even though we implemented ev$GW$ with $0 < Z < 1$ into MOLGW, we adopted ev$GW$ with $Z = 1$ in this work for a few reasons. First, Eq.~(\ref{eq:evGW.QP.eq.n}) shows that converged ev$GW$ QP energies ($\epsilon^{\text{evGW},\infty}_{m}$) is independent of whether $0 < Z < 1$ or $Z = 1$. Second, $Z = 1$ gives a unique solution that satisfies the QP equation in Eq.~(\ref{eq:evGW.QP.eq}), which allows us to avoid the $GW$ multi-solution issue from the graphical-solution and spectral-function methods and the $\sim$0.1--1~eV error from the linearization method (to be discussed in detail later). Third, ev$GW$ with $0 < Z < 1$ is suited for a simplified ev$GW$ variant that updates only a few states near HOMO and LUMO and rigidly shifts all the other states for efficiency,~\cite{Shishkin07,Blase16} but we updated all eigenvalues in this work for accuracy. Last, ev$GW$ with $Z = 1$ has no variant and does not need a QP equation solver, but ev$GW$ with $0 < Z < 1$ has multiple variants, depending on the choice of QP equation solvers. For example, two ev$GW$ variants with $0 < Z < 1$ using the linearization and graphical-solution methods in Refs.~\onlinecite{Blase11,Veril18}, respectively, may give different QP energies because the two QP equation solvers give different solutions (especially, for states far away from HOMO and LUMO).




$Z^{\text{evGW}} = 1$ and the efficiency comparison of ev$GW$ and $G_{0}W_{0}$ are discusssed in supplementary material.

\begin{table}
  \caption{Number of occupied and empty states for the $\uparrow$-spin channel ($N_{\text{occ}}^{\uparrow}$ and $N_{\text{emp}}^{\uparrow}$, respectively) used in $GW$ calculations. FC means the frozen-core approximation. AE and ECP mean all electron and effective core potential, respectively. CN means the cardinal number.}
  \label{tab:Occupied.Empty.States}
  \begin{tabular*}{0.48\textwidth}{ @{\extracolsep{\fill}} r c c c c c c c }
    \hline \hline
& & & $N_{\text{occ}}^{\uparrow}$ & \multicolumn{4}{c}{$N_{\text{emp}}^{\uparrow}$} \\
\cline{5-8}
Anion & Potential & FC & & CN=2 & CN=3 & CN=4 & CN=5 \\ \hline
ScO$^{-}$ & AE & Yes &   9 & 67 & 124 & 205 & 314 \\
ScO$^{-}$ & AE & No  & 15 & 67 & 124 & 205 & 314 \\ \hline
TiO$^{-}$ & AE & Yes & 10 & 66 & 123 & 204 & 313 \\
TiO$^{-}$ & AE & No  & 16 & 66 & 123 & 204 & 313 \\ \hline
CuO$^{-}$ & AE & Yes & 13 & 63 & 120 & 201 & 310 \\
CuO$^{-}$ & AE & No  & 19 & 63 & 120 & 201 & 310 \\
CuO$^{-}$ & ECP & Yes & 13 & 63 & 120 & 201 & 310 \\
CuO$^{-}$ & ECP & No  & 14 & 63 & 120 & 201 & 310 \\ \hline
ZnO$^{-}$ & AE & Yes & 14 & 62 & 119 & 200 & 309 \\
ZnO$^{-}$ & AE & No  & 20 & 62 & 119 & 200 & 309 \\
ZnO$^{-}$ & ECP & Yes & 14 & 62 & 119 & 200 & 309 \\
ZnO$^{-}$ & ECP & No  & 15 & 62 & 119 & 200 & 309 \\
    \hline \hline
  \end{tabular*}
\end{table}

\section{Computational Details and Test Results} \label{sec:comp.details.test.results}



\subsection{Computational Details}

Our gKS calculations were carried out using both MOLGW and NWChem in order to cross-check the results and to ascertain the correct SCF convergence. For $GW$ calculations, we used only MOLGW. MOs were expanded using augmented Dunning correlation-consistent Gaussian basis sets, aug-cc-pV$n$Z ($n$ = D, T, Q, and 5), which are designed to smoothly converge with basis size. Augmentation using diffuse functions is essential in ground-state calculations for anions and in excited-state calculations for both neutrals and anions. Without augmentation, gKS and $GW$ eigenvalues for empty states converge very slowly with basis size.~\cite{Korbel14,Wilhelm16} In the following, the cardinal number (CN = 2, 3, 4, and 5) is used to represent the approximate size of diverse basis sets employed in the literature and this work. For example, CN=4 means def2-QZVP in Ref.~\onlinecite{vanSetten15}, and aug-cc-pVQZ in this work. Table~\ref{tab:Occupied.Empty.States} summarizes the exact size of CN=2,3,4,5 basis sets used in this work. To determine the optimized bond lengths of TMO anions, we used NWChem with PBE and CN=3. We obtained bond lengths of 1.695, 1.642, 1.697, and 1.765~\AA for ScO$^-$, TiO$^-$, CuO$^-$, and ZnO$^-$, respectively.


In order to study the starting-point dependency of the $GW$ approximation, we used global hybrid functionals:
\begin{equation}
E^{\text{PBE}\alpha,\sigma}_{\text{xc}} = \alpha E^{\text{HF},\sigma}_{\text{x}} + (1 - \alpha) E^{\text{PBE},\sigma}_{\text{x}} + E^{\text{PBE},\sigma}_{\text{c}}, \label{eq:PBE.alpha}
\end{equation}
where $E^{\text{HF},\sigma}_{\text{x}}$, $E^{\text{PBE},\sigma}_{\text{x}}$, and $E^{\text{PBE},\sigma}_{\text{c}}$ are Fock exact exchange, PBE exchange, and PBE correlation energies, respectively. We refer to the hybrid functionals in Eq.~(\ref{eq:PBE.alpha}) as PBE$\alpha$ functionals in this work. While we tested other functionals such as B3LYP, HSE06, BHLYP, and HF, we discuss only PBE$\alpha$($0.00 \le \alpha \le 1.00$) results because the EXX amount in the starting point has a stronger effect on $GW$ results than other factors such as range separation (to screen the Coulomb interaction) and correlation type. As shown in supplementary material, HSE06, PBE0, and BHLYP$\alpha$($\alpha$=0.25) [where PBE is replaced by LYP in Eq.~(\ref{eq:PBE.alpha})] give similar $GW$ results. Note that the type of the correlation functional is not important (e.g. PBE vs LYP), but the existence of it is. As shown in supplementary material, PBE$\alpha$($\alpha$=1.00) and HF can make a large difference ($\sim$1~eV) in $GW$ results for some states.



\subsection{gKS Test Results}

\subsubsection{Effective Core Potentials}  \label{sec:ECP}


Unlike Sc and Ti, Cu and Zn have two choices of basis sets: AE (All Electron) and ECP (Effective Core Potential). ECP allows to remove core electrons and include relativistic effects. We first tested scalar relativistic effects by comparing AE and ECP $GW$ binding energies. We did not include spin-orbit coupling because (i) spin-orbit ECP is not implemented in MOLGW, and (ii) spin-orbit effects are very small in Cu and Zn, which are relatively light elements.~\cite{Collins95,Peterson05} The test results are presented in supplementary material. We found that (i) ECP and AE $GW$ IEs differ by 0.01--0.15~eV, depending on the subtle competition between direct and indirect relativistic effects ($s$ and $p$ orbital contraction and stabilization and $d$ and $f$ orbital expansion and destabilization, respectively),~\cite{Pyykko79,Pyykko88,Pyykko12} which is consistent with Ref.~\onlinecite{Bagus75}, and (ii) ECP $GW$ 3$d$-electron binding energies are smaller than AE ones by 0.16--0.66~eV due to indirect effects, which is consistent with Ref.~\onlinecite{Tatewaki17}. We next tested the efficiency of ECP with respect to AE. We found that ECP is more efficient than AE because the absence of core states not only makes the basis size smaller, which benefits both gKS and $GW$ parts, but also makes the SCF cycle faster and more stable, which benefits only the gKS part. In this work, we present mainly AE results not because AE is superior to ECP, but because scalar relativistic effects make smaller changes than large ($\sim$1--2~eV) errors that we encounter. However, we discuss both AE and ECP results for 3$d$-electron binding energies, where scalar relativistic effects are considerable ($\sim$10\% of the experimental 3$d$-electron binding energy). Note that Ref.~\onlinecite{vanSetten15} used only AE even though the $GW$100 benchmark set contains Ag$_{2}$, Cu$_{2}$ and NCCu molecules.

\subsubsection{RI for gKS} \label{sec:RI.gKS}

We did not use RI in this work because our goal is to assess the range of applicability of the $GW$ approximation as accurately as possible using small molecules, but RI is unavoidable for the practical $GW$ study of large molecules. Thus, we evaluated the quality of RI for both AE and ECP by comparing RI and no-RI gKS eigenvalues and total spins. The evaluation results are presented in supplementary material. We found that CN=5 RI ECP causes a large random error in gKS results (e.g. $\sim$0.2 and $\sim$0.8~eV for CuO$^-$ and ZnO$^-$, respectively, in gKS-PBE IEs), which \emph{decreases} with the EXX amount. It is important to note that unlike the SCF convergence error, which occurs only in open-shell systems with non-zero EXX amounts, this gKS RI error occurs in both closed- and open-shell systems with all EXX amounts. It is difficult to detect the gKS RI error because all SCF cycles with different convergence parameters smoothly converge to the same local minimum with no or small spin contamination. Therefore, we conclude that RI should be used only after the potential gKS RI error is thoroughly tested.


Note that because we did not use RI and the 4-center integrals are computed at each SCF step, a \emph{single} gKS calculation is as expensive as a \emph{single} $GW$ calculation in this work, which is consistent with Ref.~\onlinecite{Wilhelm16}. Note also that we discussed the effect of RI on $GW$ results in supplementary material.


\subsubsection{SCF Convergence Tests}

It is not straightforward to obtain the correct mean-field input for $GW$ calculations, because successful SCF convergence could come from both correct convergence to a global minimum as well as wrong convergence to some local minima. This is a particularly critical issue in gKS calculations on open-shell systems involving non-zero EXX and large basis. Many minima with similar total energies and total spins due to nearly degenerate energy levels in 3$d$ transtion metals make it more difficult to obtain correct SCF convergence.~\cite{Gutsev00,Gonzales00,Kudin07} For closed-shell systems or open-shell systems with (semi-)local xc functionals, the SCF cycle is generally guaranteed to converge to a global minimum. However, when EXX is used for open-shell systems, wrong SCF convergence occurs frequently and randomly, which makes manual, time-consuming, and error-prone SCF convergence tests mandatory. 


In order to obtain the correct mean-field input, we performed three-step SCF convergence tests. First, we used 12 and 96 sets of SCF convergence parameters for MOLGW and NWChem, respectively. Second, we manually searched for correctly converged SCF results using multiple indicators: gKS total energy, total spin $\langle S^{2} \rangle$ in Eq.~(\ref{eq:S2}), the number of total SCF cycles, a trend over basis size (CN=2,3,4,5), and a trend over EXX amount (by manually choosing gKS total energies and total spins that vary smoothly with basis size and EXX amount). Last, we cross-checked all MOLGW and NWChem gKS results. Our SCF convergence test results are presented in supplementary material.

Note that because of our heavy SCF convergence tests, \emph{total} gKS calculations are more expensive than \emph{total} $GW$ calculations in this work.


\subsection{$GW$ Test Results} \label{sec:GW.test.results}

\begin{figure*}
\begin{tabular}{c c}
{\includegraphics[trim=0mm 0mm 0mm 0mm, clip, width=0.48\textwidth]{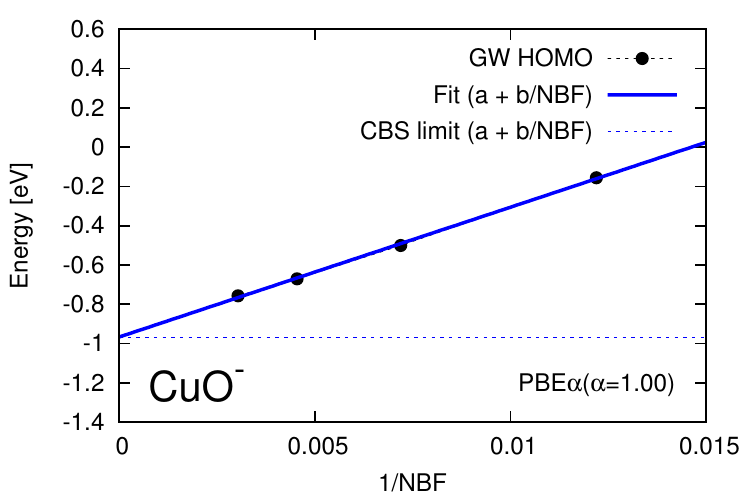}}
{\includegraphics[trim=0mm 0mm 0mm 0mm, clip, width=0.48\textwidth]{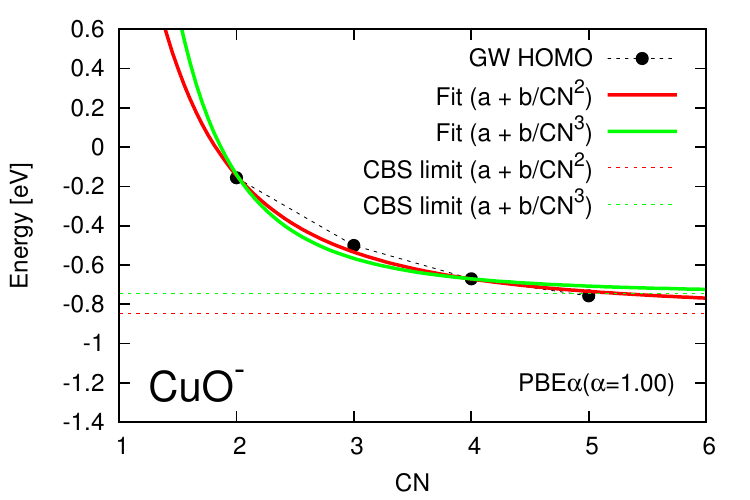}}
\end{tabular}
\caption{(Color online) Effect of the fitting function type on the $GW$ complete basis set (CBS) limit. The calculations are presented for the HOMO of CuO$^{-}$ at the $G_{0}W_{0}@\text{PBE}\alpha$($\alpha$=1.00) level of theory. NBF and CN represent the number of basis functions and the cardinal number, respectively. $G_{0}W_{0}@\text{PBE}\alpha$($\alpha$=1.00) HOMO energies are obtained from gKS-PBE$\alpha$($\alpha$=1.00) HOMO-1, which corresponds to gKS-PBE HOMO (see text).}
\label{fit.NBF.CN23.HOMO.0.EXX100}
\end{figure*}

\begin{figure*}
\begin{tabular}{c c}
{\includegraphics[trim=0mm 0mm 0mm 0mm, clip, width=0.48\textwidth]{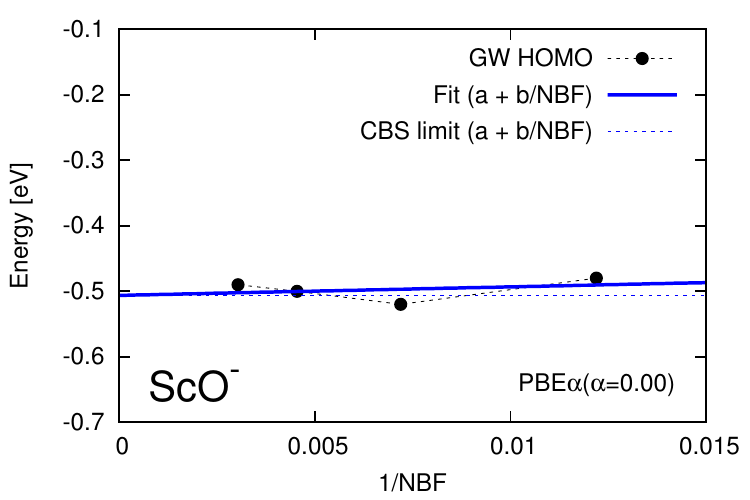}}
{\includegraphics[trim=0mm 0mm 0mm 0mm, clip, width=0.48\textwidth]{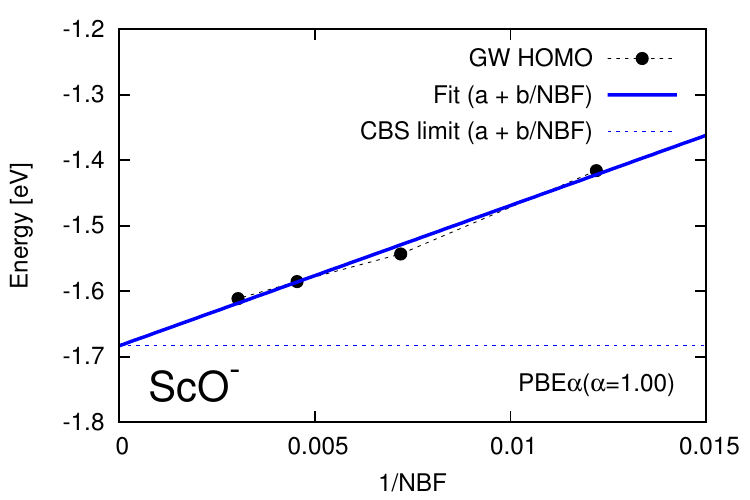}}
\end{tabular}
\caption{(Color online) Effect of the EXX amount on the $GW$ complete basis set (CBS) limit. The calculations are presented for the HOMO of ScO$^-$ at the $G_0W_0@\text{PBE}$ (left) and $G_0W_0@\text{PBE}\alpha$($\alpha$=1.00) (right) levels of theory. NBF represents the number of basis functions.}
\label{fit.NBF.HOMO.0.EXX000.EXX100}
\end{figure*}

\subsubsection{Complete Basis Set Limit} \label{sec:CBS}

Like MP2, RPA, and CCSD(T) correlation energies, $GW$ QP energies converge slowly with basis size. Accordingly, one should extrapolate $GW$ QP energies obtained from different basis sizes to the complete basis set (CBS) limit to avoid the incomplete basis set error of $\sim$0.1~eV.~\cite{Ren12} We, therefore, tested the effect of fitting function type, EXX amount, and basis size on the $GW$ CBS limit.


Two fitting functions are most widely used for the CBS limit,~\cite{Hattig12} which we refer to as standard fitting functions in this work:
\begin{align}
E_{m} &= a + \frac {b} {N_{\text{BF}}}, \label{eq:fit.NBF} \\
E_{m} &= a + \frac {b} {{\rm CN}^{3}}, \label{eq:fit.CN3}
\end{align}
where $E_{m}$ are correlation or $m$th QP energies, $a$ and $b$ are fitting parameters, $N_{\text{BF}}$ is the number of basis functions (see Table~\ref{tab:Occupied.Empty.States}), and CN is the cardinal number. In Eqs.~(\ref{eq:fit.NBF}) and (\ref{eq:fit.CN3}), $a$ gives the correlation or QP energy in the CBS limit. Note that there are various non-standard fitting functions used in the literature.~\cite{Huser13,Rangel16,Bruneval16,Hung17a,Shi18}

Fig.~\ref{fit.NBF.CN23.HOMO.0.EXX100} compares CBS results obtained from two standard fitting functions in Eqs.~(\ref{eq:fit.NBF}) and (\ref{eq:fit.CN3}) (as well as one non-standard one used in Refs.~\onlinecite{Bruneval16,Shi18}). We see that different fitting functions always give different $GW$ CBS limits, deviating from each other by up to $\sim$0.1~eV depending on molecular systems and molecular orbitals. Fig.~\ref{fit.NBF.HOMO.0.EXX000.EXX100} shows the effect of the EXX amount on the $GW$ CBS limit. We observe that the incomplete basis set error increases with the EXX amount. CN=2 occasionally and randomly causes a significant error ($\sim$0.1~eV) in the $GW$ CBS limit, which is commonly observed in the literature.~\cite{vanSetten13,vanSetten15} Based on these test results, we conclude that it is important to check whether extrapolation is used or not, whether CN=2 is used or not for extrapolation, and which fitting function is used when analyzing and comparing $GW$ results. For example, Ref.~\onlinecite{vanSetten15} reported that IEs obtained from Gaussian- and planewave (PW)-based $GW$ implementations with and without extrapolation, respectively, differ by $\sim$0.2~eV, but Refs.~\onlinecite{Maggio17,Govoni18} showed that the use of PW $GW$ IEs with extrapolation reduces the difference to $\sim$0.06~eV.

In this work, we obtained gKS and $GW$ CBS results using the fitting function in Eq.~(\ref{eq:fit.NBF}) with CN=2,3,4,5. We chose Eq.~(\ref{eq:fit.NBF}) not because it is superior to Eq.~(\ref{eq:fit.CN3}), but because it can be used by both Gaussian- and PW-based $GW$ implementations.



\subsubsection{Number of Empty States}

By enabling MOLGW 1.F to support the largest available basis set (CN=5), we also tested the effect of CN=5 on the $GW$ CBS limit. The test results are presented in supplementary material. Here, we briefly mention a couple of trends. In most cases, CN=5 has a small ($\sim$10~meV) effect on the $GW$ CBS limit, since CN=4,5 $GW$ QP energies are very similar. However, in some cases, CN=5 has an appreciable ($\sim$0.1~eV) effect on the $GW$ CBS limit by reducing the effect of the large random CN=2 error on the $GW$ CBS limit. In other words, CN=5 barely improves the accuracy of the $GW$ CBS limit, but mostly acts as a bumper for the CN=2 error. Moreover, CN=5 calculations are expensive (due to the large number of empty states and the slow SCF convergence speed) and error-prone (due to the high chance of SCF convergence and gKS RI errors). Therefore, we conclude that it is more beneficial to obtain the $GW$ CBS limit from CN=3,4 than from CN=2,3,4,5. Using CN=4 ($\sim$100 empty states per atom, as shown in Table~\ref{tab:Occupied.Empty.States}) instead of CN=5 as the largest basis set for the $GW$ CBS limit tremendously reduces the computational costs. This conclusion is consistent with Ref.~\onlinecite{vanSetten15}, which used only CN=3,4 for extrapolation, and gives a rough estimate for two important and inter-dependent convergence parameters for $\Sigma^{\sigma}_{\text{c}}(\omega)$, the dimension of the dielectrix matrix and the number of empty states, in sum-over-states PW $GW$ implementations.~\cite{vanSetten15}


The above conclusion holds only for occupied states. The effect of CN=5 on the $GW$ CBS limit for empty states is discussed in supplementary material. The effect of the number of occupied states on $GW$ results using the frozen-core (FC) approximation, which reduces the number of occupied states used in the construction of $G$ and $W$ and thus speeds up $GW$ calculations,~\cite{Bruneval13} is also discussed in supplementary material.

\subsubsection{$G_{0}W_{0}$ Quasiparticle Energy} \label{sec:test.G0W0.QP.energy}

A full-frequency $G_{0}W_{0}$ method used in this work produces complicated self-energy pole (and spectral-function peak) structures at non-frontier orbitals, so it is not straightforward to automatically obtain correct and accurate $\epsilon^{\text{G}_{0}\text{W}_{0}}_{m}$ by using a single value of $\eta$ and a single QP equation solver. Thus, we used three values of $\eta$ (0.001, 0.002, and 0.005~Ha with $\Delta\omega$ = 0.001~Ha) and three QP equation solvers mentioned in Section~\ref{sec:background.G0W0.QP.energy} (linearization, graphical-solution, and spectral-function methods).

Our analysis of a total of nine solutions for $\epsilon^{\text{G}_{0}\text{W}_{0}}_{m}$ shows that the graphical-solution method using $\eta$ = 0.001~Ha and the linearization method randomly give incorrect solutions. Therefore, we obtained $\epsilon^{\text{G}_{0}\text{W}_{0}}_{m}$ from the graphical-solution or spectral-function method using $\eta$ = 0.002 or 0.005~Ha. When $\epsilon^{\text{G}_{0}\text{W}_{0},\text{graph}}_{m}$ and $\epsilon^{\text{G}_{0}\text{W}_{0},\text{spect}}_{m}$ are different, we manually selected a correct solution by analyzing $\Sigma_{\text{c}}(\omega)$ and $A(\omega)$.

\begin{figure*}
\begin{tabular}{c c}
{\includegraphics[trim=0mm 7mm 0mm 0mm, clip, width=0.48\textwidth]{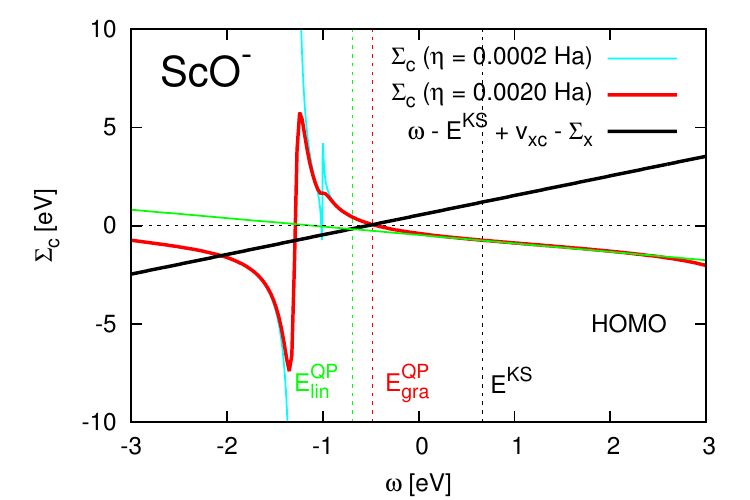}}
{\includegraphics[trim=0mm 7mm 0mm 0mm, clip, width=0.48\textwidth]{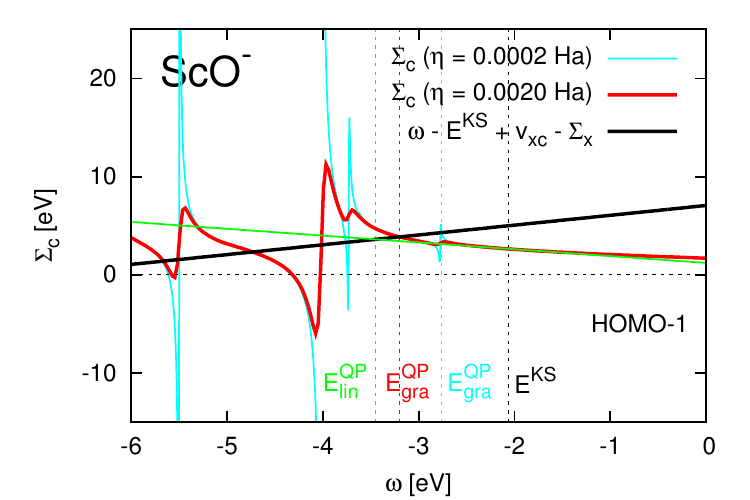}} \\
{\includegraphics[trim=0mm 0mm 0mm 0mm, clip, width=0.48\textwidth]{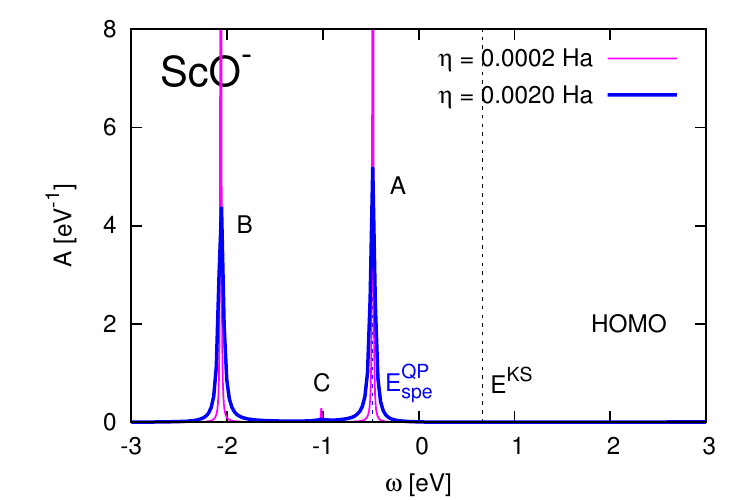}}
{\includegraphics[trim=0mm 0mm 0mm 0mm, clip, width=0.48\textwidth]{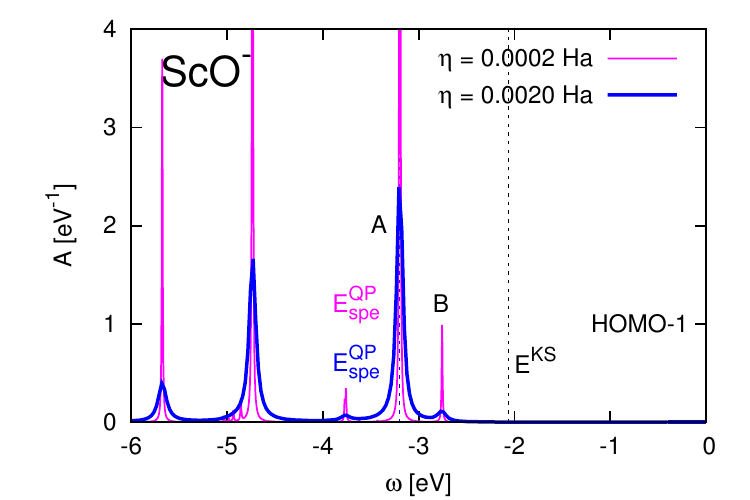}} \\
{\includegraphics[trim=0mm 7mm 0mm 0mm, clip, width=0.48\textwidth]{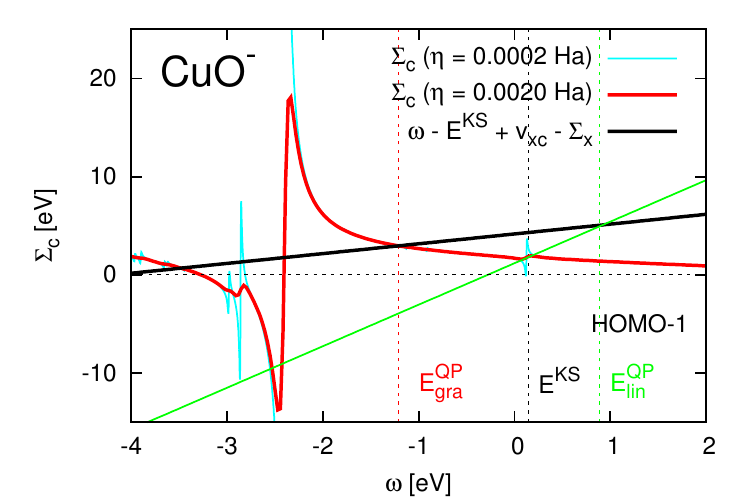}}
{\includegraphics[trim=0mm 7mm 0mm 0mm, clip, width=0.48\textwidth]{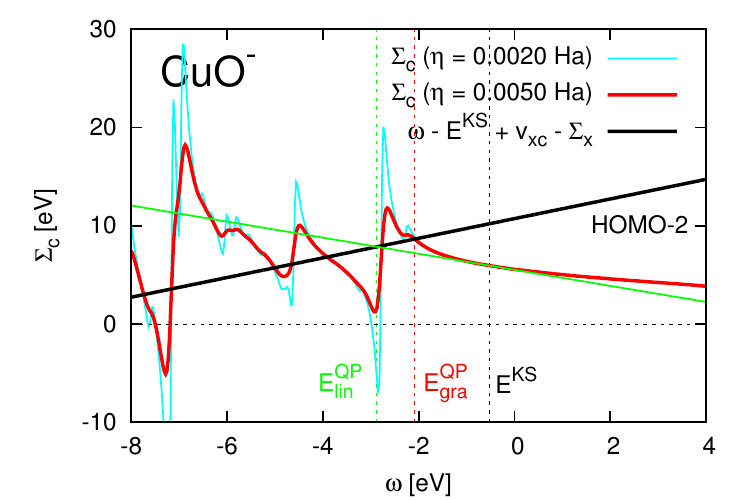}} \\
{\includegraphics[trim=0mm 0mm 0mm 0mm, clip, width=0.48\textwidth]{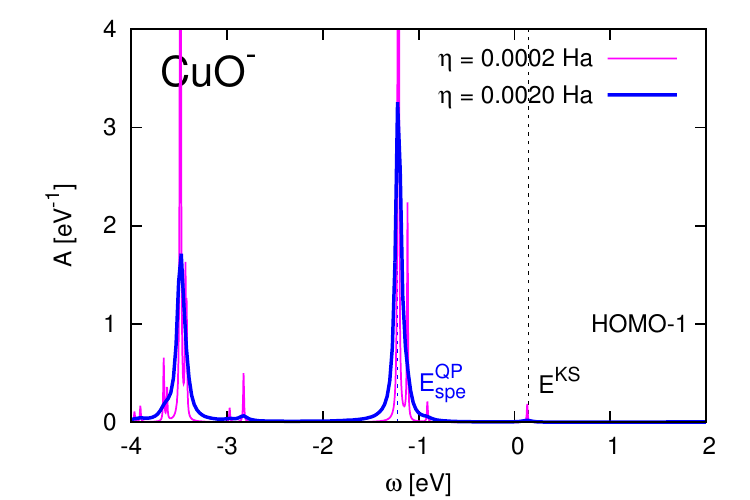}}
{\includegraphics[trim=0mm 0mm 0mm 0mm, clip, width=0.48\textwidth]{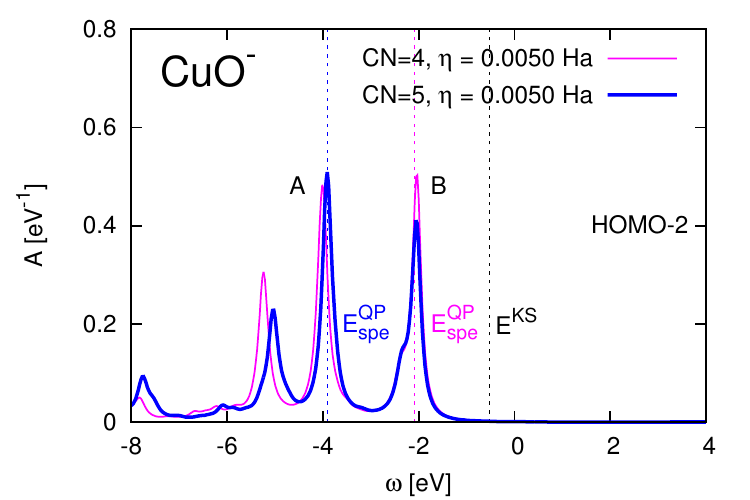}}
\end{tabular}
\caption{(Color online) Comparison of three QP equation solvers: linearization, graphical-solution, and spectral-function methods. (Top left) All the three methods give correct solutions. (Top right) The graphical-solution method can give incorrect solutions. (Bottom left) The linearization method can give incorrect solutions. (Bottom right) The spectral-function method can give incorrect solutions. $E^{\text{KS}}$ represents the gKS-PBE eigenvalue. $E^{\text{QP}}_{\text{lin}}$, $E^{\text{QP}}_{\text{gra}}$, and $E^{\text{QP}}_{\text{spe}}$ represent $G_{0}W_{0}@\text{PBE}$ QP energies obtained from linearization, graphical-solution, and spectral-function methods, respectively. The green straight line is a tangent to the red curve at $\omega = E^{\text{KS}}$. Except for $A(\omega)$ at the bottom right, all results are obtained from CN=2 no-RI AE.}
\label{Zqp.same.diff}
\end{figure*}

A large distance between $\epsilon_{m}$ and $\epsilon^{\text{G}_{0}\text{W}_{0}}_{m}$ and multiple self-energy poles between them makes it difficult to obtain correct and accurate $\epsilon^{\text{G}_{0}\text{W}_{0}}_{m}$ for non-frontier orbitals, which is especially the case for (semi-)local xc functionals. Fig.~\ref{Zqp.same.diff} shows various examples of successes and failures of three QP equation solvers for $G_{0}W_{0}@\text{PBE}$. In the following, we analyze each example individually. Note that we used $\eta$ = 0.0001, 0.0002, and 0.0005~Ha with $\Delta\omega$ = 0.0001~Ha in a few of the following examples to demonstrate the danger of small $\eta$ values for the spectral-function method. Such small $\eta$ values are not recommended, as they significantly increase disk storage requirements while barely improving accuracy.


First, the top two left panels of Fig.~\ref{Zqp.same.diff} show a general example, in which all three methods succeed. We see a few general trends. Typically, all three methods give correct solutions at $m$ = HOMO and LUMO, which have a simple pole structure in $\Sigma_{\text{c}}(\omega)$. Graphical-solution and spectral-function methods generally give \emph{multiple} solutions, whereas the linearization method always gives a \emph{unique} solution, at which two straight lines intersect. Generally, graphical-solution and spectral-function methods give identical (correct and accurate) solutions (in this case, at $\omega = -0.48$~eV), whereas the linearization method gives a different (correct but inaccurate) solution (in this case, at $\omega = -0.69$~eV) due to an intrinsic error of $\sim$0.1--1~eV. A very small $\eta$ (0.0002~Ha) sharpens a weak self-energy pole at $\omega = -1.0$~eV in $\Sigma_{\text{c}}(\omega)$, but the sharpened pole \emph{does not} cause an error in the graphical-solution method because it is not between $\epsilon^{\text{G}_{0}\text{W}_{0}}_{m}$ and $\epsilon_{m}$. The very small $\eta$ heightens the weak peak C in $A(\omega)$, but the heightened peak \emph{cannot} cause an error in the spectral-function method because it is still lower than other peaks A and B. In other words, \emph{the spectral-function method depends more weakly on the choice of $\eta$ than the graphical-solution method.} The very small $\eta$ has little effect on the linearization method, because (i) the linearization method in this work depends only on $\Sigma_{\text{c}}(\epsilon_{m} \pm \Delta\omega)$, and (ii) $\omega = -1.0$~eV is too distant from $\epsilon_{m}$ to affect the finite difference method.

Second, the top two right panels of Fig.~\ref{Zqp.same.diff} show a special example, in which the graphical-solution method can give an incorrect solution. We see a few special trends. Generally, some of the three methods give incorrect solutions at $m$ = HOMO-$n$ and LUMO+$n$ ($n$ = 1, 2, 3, ...), which have a complicated pole structure in $\Sigma_{\text{c}}(\omega)$. A very small $\eta$ (0.0002~Ha) sharpens a weak pole at $\omega = -2.8$~eV in $\Sigma_{\text{c}}(\omega)$, and this sharpened pole causes a large error of 0.4~eV in the graphical-solution method because it is between $\epsilon^{\text{G}_{0}\text{W}_{0}}_{m}$ and $\epsilon_{m}$ and sharpened enough to make the secant method fail by causing it to find an incorrect intersection point. The very small $\eta$ heightens a weak peak B in $A(\omega)$, but the heightened peak \emph{cannot} cause an error in the spectral-function method because it is still lower than the other peak A. The very small $\eta$ has little effect on the linearization method because $\omega = -2.8$~eV is distant from $\epsilon_{m}$.

Third, the bottom two left panels of Fig.~\ref{Zqp.same.diff} show a special example, in which the linearization method can give an incorrect solution. We see a few special trends. A very small $\eta$ (0.0002~Ha) sharpens a weak pole at $\omega \approx \epsilon_{m}$ in $\Sigma_{\text{c}}(\omega)$, but the sharpened pole \emph{does not} cause an error in the graphical-solution method even though it is between $\epsilon^{\text{G}_{0}\text{W}_{0}}_{m}$ and $\epsilon_{m}$, because the pole is not sharpened enough ($\eta$ = 0.0001~Ha, on the other hand, causes a large error of 1.3~eV). The very small $\eta$ heightens a weak peak at $\omega \approx \epsilon_{m}$ in $A(\omega)$, but the heightened peak \emph{cannot} cause an error in the spectral-function method because it is still lower than the other peak at $\omega \approx \epsilon^{\text{G}_{0}\text{W}_{0}}_{m}$. The very small $\eta$ sharpens a weak pole at $\omega \approx \epsilon_{m}$ in $\Sigma_{\text{c}}(\omega)$, and the sharpened pole causes a large error of 2.1~eV in the linearization method because it is too close to $\epsilon_{m}$, making the finite difference method fail by causing a large error in the slope of the tangent line at $\omega = \epsilon_{m}$.

Last, the bottom two right panels of Fig.~\ref{Zqp.same.diff} show a special example, in which the spectral-function method can give an incorrect solution. We see a few special trends. A very small $\eta$ (0.0002~Ha) sharpens a weak pole at $\omega = -2.3$~eV in $\Sigma_{\text{c}}(\omega)$, but the sharpened pole \emph{does not} cause an error in the graphical-solution method because it is not between $\epsilon^{\text{G}_{0}\text{W}_{0}}_{m}$ and $\epsilon_{m}$. Two peaks A and B (at $\omega = -3.9$ and $-2.1$~eV, respectively) in $A(\omega)$ have similar spectral weights (and peak heights), so it is not straightforward to unambiguously determine which one is a QP peak or a satellite. Spectral weights (practically, peak heights) of the two peaks depend on the basis size: the peak B (A) is higher than the peak A (B) for CN=2,3,4 (CN=5). We \emph{chose} peak B as the QP peak, because (i) it is consistent with the solution from the graphical-solution method, and (ii) it is consistent with a trend over EXX amount [$G_{0}W_{0}@\text{PBE}\alpha(0.00 \le \alpha \le 1.00)$ QP HOMO-2 energies of CuO$^{-}$ using the solution from the peak B vary smoothly with $\alpha$, as shown in Fig.~\ref{G0W0.EXX.CuO.ZnO}]. The choice of $\eta$ and CN has little effect on the linearization method, but $\epsilon^{\text{G}_{0}\text{W}_{0},\text{linear}}_{m} = -2.9$~eV causes a large overestimation error of 0.8~eV in $G_{0}W_{0}@\text{PBE}$ binding energy.


There are several points to note about the above examples: (i) we chose simple examples, in which only one method can give an incorrect solution, for demonstration purposes; multiple methods can give incorrect solutions simultaneously, as shown in supplementary material, (ii) not only a very small $\eta$, but also a very large $\eta$ (e.g. $\sim$0.05~Ha in Ref.~\onlinecite{Hung17b}) can cause a large error, (iii) deep states (e.g. HOMO-$n$, where $n$ = 5, 6, ...) have much more complicated pole [peak] structures in $\Sigma_{\text{c}}(\omega)$ [$A(\omega)$] than those in Fig.~\ref{Zqp.same.diff}, so it is very difficult to choose correct and accurate $\epsilon^{\text{G}_{0}\text{W}_{0}}_{m}$ for deep states not only automatically, but also manually.


We conclude this section by summarizing several guidelines to obtain a reliable and reproducible $G_{0}W_{0}@\text{PBE}$ QP spectrum. First, one should try multiple $\eta$ (and $\Delta\omega$) values. There is no single general $\eta$ value that works well for all QP equation solvers, molecular systems, and molecular orbitals. In other words, while $\eta$ is typically viewed as a convergence parameter (the smaller $\eta$, the more accurate $GW$ QP energy), it is practically an adjustable parameter, which should be not too small or too large (e.g. 0 and $\sim$0.05~Ha in Refs.~\onlinecite{Veril18,Hung17b}, respectively). The optimal value of $\eta$ depends on $|\epsilon^{\text{G}_{0}\text{W}_{0}}_{m}$ - $\epsilon_{m}|$, which generally decreases with the amount of EXX and increases with the depth of the $m$th state. For example, when calculating $G_{0}W_{0}@\text{PBE}$ HOMO and LOMO (lowest occupied molecular orbital) energies, one may try $\sim$0.1 and $\sim$1~eV, respectively, for $\eta$.


Second, we recommend using multiple QP equation solvers. As shown in Fig.~\ref{Zqp.same.diff}, the $G_{0}W_{0}@\text{PBE}$ QP spectrum automatically obtained from a single QP equation solver can contain a large ($\sim$1~eV) error at random states.


Third, we recommend using multiple basis sizes. As shown in the bottom right of Fig.~\ref{Zqp.same.diff}, different basis sets with different sizes can give very different $G_{0}W_{0}$ QP energies (by $\sim$1~eV) at random states. Using multiple basis sizes allows for not only accurate $GW$ results without small ($\sim$0.1~eV) systematic errors from the basis set incompleteness, but also correct gKS and $GW$ results without large ($\sim$1~eV) random errors from SCF convergence and $GW$ multi-solution issues, respectively.


Fourth, one should be fully aware of the large random errors that the linearization method, which is the most widely used QP equation solver, can cause. Ref.~\onlinecite{Maggio17} suggests the linearization method as a preferable method for a fair comparison of $G_{0}W_{0}@\text{PBE}$ IE (and EA) from different $GW$ implementations, because it gives a unique solution and thus is free of the $GW$ multi-solution issue. The idea works well for the IE, but it does not perform as well for the QP spectrum. For HOMO (and LUMO), the linearization method generally succeeds and systemically overestimates the IE only by $\sim$0.1~eV with respect to the accurate one from the graphical-solution and spectral-function methods, as shown in the top left of Fig.~\ref{Zqp.same.diff}, accidentally reducing the $\sim$0.5~eV underestimation error by $G_{0}W_{0}@\text{PBE}$ with respect to experiment.~\cite{vanSetten15,Maggio17,Govoni18} However, for deep states, it randomly succeeds or fails, as shown in the bottom left of Fig.~\ref{Zqp.same.diff}, and randomly overestimates or underestimates $G_{0}W_{0}@\text{PBE}$ binding energies by $\sim$1~eV compared to accurate ones, as shown in the bottom right of Fig.~\ref{Zqp.same.diff} and supplementary material, respectively. This large and unpredictable (with respect to state, magnitude, and direction) error makes the linearization method inadequate for the $G_{0}W_{0}@\text{PBE}$ QP spectrum.


Last, one should be aware that different ways to handle the $GW$ multi-solution issue are found in the literature. For example, Ref.~\onlinecite{vanSetten15}, which suggests that all solutions are physically relevant, manually searched for an actually relevant one by varying $\eta$, whereas Ref.~\onlinecite{Veril18} avoided the issue by automatically selecting the solution with the largest $Z_{m}$ in Eq.~(\ref{eq:Z.GW100}) and using only $\eta = 0$. In this work, we adopted a combined approach. In other words, we automatically chose the solution with the highest spectral weight, which is identical to the solution with the largest $Z_{m}$, as explained in Section~\ref{sec:background.G0W0.QP.energy}, when one solution is clearly more relevant than others, but we manually selected one solution that gives smoothly varying $GW$ binding energies with a change in $G_{0}W_{0}$ starting point and ev$GW$ self-consistency level without causing unphysical kinks when multiple solutions are equally relevant [e.g. two solutions at peaks A and B in the bottom two right panels of Fig.~\ref{Zqp.same.diff} give similar $Z_{m}$ ($\sim$0.2) for $m = \text{HOMO-2}$ of CuO$^{-}$ due to similar slopes of the self-energy (approximately, $-4.0$)]. However, when it comes to $\eta$, we adopted the approach of Ref.~\onlinecite{vanSetten15} instead of that of Ref.~\onlinecite{Veril18} because $\eta = 0$ frequently causes the secant method in the graphical-solution method to find an incorrect intersection point and makes $\langle \varphi^{\sigma}_{m} | \Sigma^{\sigma}_{\text{c}} (\omega) | \varphi^{\sigma}_{m} \rangle$ in Eq.~(\ref{eq:Sigma.c}) diverge at $\omega = \epsilon^{\sigma}_{i} - \Omega_{s}$ and $\omega = \epsilon^{\sigma}_{a} + \Omega_{s}$. The cumulant expansion,~\cite{Hedin99,Zhou15} which describes plasmon satellites better than the $GW$ approximation, may allow us to address the $GW$ multi-solution issue when the QP picture breaks down, but it is beyond the scope of this work.


\subsubsection{$G_{n}W_{0}$ and $G_{n}W_{n}$ Quasiparticle Energy} \label{sec:test.evGW.QP.energy}

In this work, we used only $\eta = 0.001$~Ha for ev$GW$ because it is small enough to obtain the convergence of ev$GW$ QP energies with respect to $\eta$ within $\sim$0.01~eV. The convergence test results for ev$GW$ QP energies with respect to the iteration number are shown in supplementary material. QS$GW$ and our ev$GW$ are quasiparticle-only $GW$ schemes with no spectral weight transfer ($Z = 1$), and $G_{n}W_{n}$ is a diagonal approximation to QS$GW$. Therefore, we compared the convergence behaviors of QS$GW$ and our ev$GW$ and found a couple of similarities and differences between them.


First, the ev$GW$ convergence is reached after only a few iterations, which is consistent with the literature.~\cite{Shishkin07,Rangel16,Blase16} Due to the fast and stable convergence, a mixing scheme is not used in our ev$GW$. Unlike ev$GW$, QS$GW$ generally needs $\sim$10--20 (up to 60) iterations and a mixing scheme.~\cite{Bruneval12,Koval14,Kaplan16}


Second, the orbital character affects the starting-point dependency of ev$GW$. For example, we observed that ev$GW$ QP energies for HOMO of CuO$^{-}$ depend more strongly on the EXX amount in the ev$GW$ starting point than those for HOMO of ScO$^{-}$. We attribute this to different amounts of 3$d$ character in HOMOs of ScO$^{-}$ and CuO$^{-}$ (6\% and 23\%, respectively, as shown in Table~\ref{tab:Orbital.Analysis}). In other words, as the 3$d$ character in MO increases, the starting-point dependency of ev$GW$ increases. We also observed that $G_{n}W_{n}$ has a weaker (stronger) starting-point dependency for HOMO of ScO$^{-}$ (CuO$^{-}$) than $G_{n}W_{0}$. Our observations for ScO$^{-}$ are consistent with Ref.~\onlinecite{Blase16}, which studied the ev$GW$ starting-point dependency using small water clusters and concluded that as the ev$GW$ self-consistency level increases from $G_{n}W_{0}$ and $G_{n}W_{n}$, the ev$GW$ starting-point dependency decreases. However, our observations for CuO$^{-}$ are not consistent with this conclusion. This is likely because CuO$^{-}$ has strong 3$d$ character in HOMO, whereas ScO$^{-}$ and small water clusters do not. This orbital-character-dependent starting-point dependency of ev$GW$ may be related to conflicting results for QS$GW$ in the literature: Ref.~\onlinecite{Koval14} showed the starting-point independency of QS$GW$ using a small $sp-$bonded molecule (CH$_{4}$), while Ref.~\onlinecite{Liao11} showed the strong starting-point dependency of QS$GW$ using a $d$ solid ($\alpha$-Fe$_{2}$O$_{3}$).


\begin{table}
  \caption{TM 3$d$ character in molecular orbitals of TMO anions, obtained from NWChem gKS-PBE and gKS-PBE$\alpha$($\alpha$=1.00) results with CN=2 no-RI AE using the Mulliken population analysis. The gKS-PBE orbital order is used for gKS-PBE$\alpha$($\alpha$=1.00) molecular orbitals (see text). Bold numbers are used to highlight entire TM 3$d$ character.}
  \label{tab:Orbital.Analysis}
  \begin{tabular*}{0.48\textwidth}{ @{\extracolsep{\fill}} l c c c c c c }
    \hline \hline
& ScO$^{-}$ & \multicolumn{2}{c}{TiO$^{-}$} & CuO$^{-}$ & \multicolumn{2}{c}{ZnO$^{-}$} \\
\cline{3-4}
\cline{6-7}
& & $\uparrow$ & $\downarrow$ & & $\uparrow$ & $\downarrow$ \\ \hline
PBE & & & & & & \\
HOMO    & 0.06	& \textbf{1.00} & 0.00 & 0.23	            & 0.00 & 0.00 \\
HOMO-1 & 0.20	& 0.09             & 0.17 & 0.32	            & 0.00 & 0.00 \\
HOMO-2 & 0.18	& 0.21             & 0.20 & \textbf{1.00} 	& 0.05 &        \\
HOMO-3 &        	& 0.24             &         & 0.75             	&        &        \\
HOMO-4 &        	&                    &         & 0.48             	&        &        \\
\hline
PBE$\alpha$($\alpha$=1.00)  & & & & & & \\
HOMO    & 0.00	& \textbf{1.00} & 0.00 & 0.10	& 0.00 & 0.00 \\
HOMO-1 & 0.13	& 0.07             & 0.10 & 0.05	& 0.00 & 0.00 \\
HOMO-2 & 0.15	& 0.14             & 0.14 & \textbf{1.00} 	& 0.06 &        \\
HOMO-3 &        	& 0.22             &         & 0.94             	&        &        \\
HOMO-4 &        	&                    &         & 0.80             	&        &        \\
    \hline \hline
  \end{tabular*}
\end{table}


\begin{figure}
\begin{tabular}{c}
{\includegraphics[trim=0mm 0mm 45mm 0mm, clip, width=0.50\textwidth]{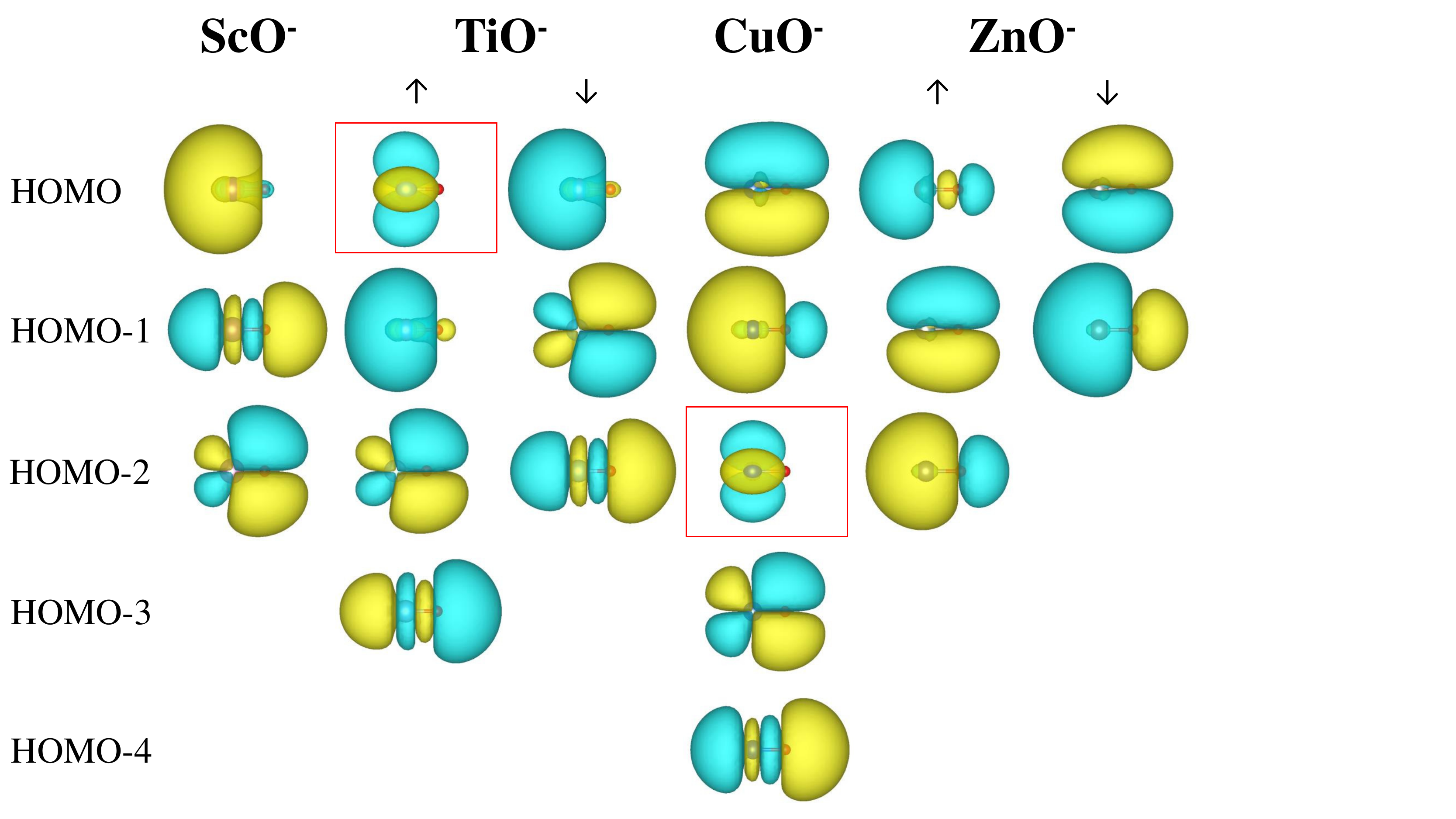}}
\end{tabular}
\caption{(Color online) Contour plots of molecular orbitals of TMO anions, obtained from MOLGW gKS-PBE results with CN=2 no-RI AE using VESTA~\cite{Momma08}. Red boxes are used to highlight entire TM 3$d$ character.}
\label{Orbital.Plot}
\end{figure}


\begin{table*}
\caption{Experimental (PES) and calculated (AE $GW$) electron binding energies of TMO anions (in eV). MAE represents the mean absolute error. Bold numbers are used to highlight 3$d$-electron binding energies.}
\label{tab:BE.exp.ours.others}
\begin{tabular*}{1.00\textwidth}{ @{\extracolsep{\fill}} l l c c c c c c c c l}
\hline \hline
& & & \multicolumn{5}{c}{$G_{0}W_{0}@\text{PBE}\alpha$} & $G_{n}W_{0}@$ & $G_{n}W_{n}@$ & \\
\cline{4-8}
& State & Exp. & $\alpha$=0.00 & $\alpha$=0.25 & $\alpha$=0.50 & $\alpha$=0.75 & $\alpha$=1.00 & PBE & PBE & Others \\ \hline
ScO$^{-}$ ($^{1}\Sigma^{+}$)	&						&		& 		& 		& 		& 		& 		& 		& 		&	\\
HOMO    						& $^{1}\Sigma^{+}$	& 1.35\footnotemark[1]	& 0.51 	& 1.15 	& 1.45	& 1.59	& 1.63 	& 1.21	& 1.35	& 1.28\footnotemark[7], 1.26\footnotemark[10], 1.19\footnotemark[11]	\\
HOMO-1 						& $^{2}\Delta$			& 3.10\footnotemark[1]	& 3.30 	& 4.77 	& 5.45	& 5.72	& 5.82 	& 5.34	& 6.08	& 2.41\footnotemark[7], 2.78\footnotemark[8], 3.31\footnotemark[9]	\\
HOMO-2 						& $^{2}\Pi$ 				& 3.40\footnotemark[1]	& 3.42  	& 4.81 	& 5.40	& 5.61	& 5.63	& 5.39	& 6.17	& 3.34\footnotemark[7], 3.24\footnotemark[8], 3.44\footnotemark[9]	\\
MAE							&						&		& 0.84\footnotemark[5]	& 0.18\footnotemark[5]	& 0.10\footnotemark[5]	& 0.24\footnotemark[5]	& 0.28\footnotemark[5]	& 0.14\footnotemark[5]	& 0.00\footnotemark[5]	&	\\
\hline
TiO$^{-}$ ($^{2}\Delta$)		&						& 		& 		& 		& 		& 		& 		& 		&		&	\\
$\uparrow$-HOMO    				& $^{1}\Sigma^{+}$	& \textbf{2.00}\footnotemark[2]	& \textbf{0.26} 	& \textbf{2.24} 	& \textbf{3.70}	& \textbf{4.79}	& \textbf{5.63} 	& \textbf{1.83}	& \textbf{2.74}	& 2.39\footnotemark[7], 2.37\footnotemark[8], 2.34\footnotemark[9]	\\
$\uparrow$-HOMO-1 				& $^{1}\Delta$			& 1.73\footnotemark[2]	& 0.53 	& 1.28 	& 1.65	& 1.83	& 1.95 	& 1.38	& 1.61	& 1.88\footnotemark[7], 1.72\footnotemark[9]	\\
$\downarrow$-HOMO				& $^{3}\Delta$			& 1.30\footnotemark[2]	& 0.31  	& 1.00 	& 1.29	& 1.43	& 1.55	& 1.06	& 1.21	& 1.19\footnotemark[7], 1.18\footnotemark[10], 1.14\footnotemark[14]	\\
MAE							&						&		& 1.31	& 0.33	& 0.60	& 1.01	& 1.37	& 0.25	& 0.32	&	\\
\hline
CuO$^{-}$ ($^{1}\Sigma^{+}$)	&						&		& 		& 		& 		& 		& 		& 		& 		&	\\
HOMO    						& $^{2}\Pi$				& 1.78\footnotemark[3]	& 0.40 	& 1.40 	& 1.58	& 1.40	& 0.97 	& 2.19	& 3.17	& 1.55\footnotemark[7], 1.52\footnotemark[10], 0.46\footnotemark[15]	\\
HOMO-1 						& $^{2}\Sigma^{+}$	& 2.75\footnotemark[3]	& 1.39 	& 2.17 	& 2.23	& 1.98	& 1.56 	& 2.66	& 3.58 & 2.96\footnotemark[7], 2.86\footnotemark[8], 1.60\footnotemark[15], 2.78\footnotemark[16], 2.47\footnotemark[17], 2.81\footnotemark[18]		\\
HOMO-2 						&						& \textbf{4.50}\footnotemark[6]	& \textbf{2.18}  	& \textbf{4.05} 	& \textbf{4.60}	& \textbf{4.56}	& \textbf{4.25}	& \textbf{4.88}	& \textbf{6.38}	& 4.07\footnotemark[7], 4.01\footnotemark[8], 4.58\footnotemark[16], 4.50\footnotemark[17]	\\
HOMO-3 						&						&		& 2.89  	& 4.49 	& 5.06	& 5.04	& 4.67	& 4.96	& 6.60	&	\\
HOMO-4 						&						&		& 3.70  	& 4.49 	& 4.63	& 4.53	& 4.24	& 5.16	& 6.39	&	\\
MAE							&						&		& 1.69	& 0.47	& 0.27	& 0.40	& 0.75	& 0.29	& 1.37	&	\\
\hline
ZnO$^{-}$ ($^{2}\Sigma^{+}$) 	& 						&		& 		& 		& 		& 		& 		& 		& 		&	\\
$\uparrow$-HOMO    				& $^{1}\Sigma^{+}$	& 2.09\footnotemark[4]	& 0.91 	& 1.91 	& 2.20	& 2.23	& 2.00 	& 2.11	& 2.57	& 2.19\footnotemark[7], 2.33\footnotemark[10], 2.29\footnotemark[12], 2.10\footnotemark[13], 1.06\footnotemark[15]	\\
$\uparrow$-HOMO-1 				& $^{1}\Pi$				& 2.71\footnotemark[4]	& 1.36 	& 2.43 	& 3.03	& 3.72	& 4.79 	& 2.77	& 3.73	& 2.62\footnotemark[7], 1.43\footnotemark[15]	\\
$\uparrow$-HOMO-2 				&						&		& 3.24 	& 4.04 	& 4.89	& 5.64	& 6.90 	& 4.77	& 5.66	& 3.50\footnotemark[15]	\\
$\downarrow$-HOMO				& $^{3}\Pi$				& 2.40\footnotemark[4]	& 1.17  	& 2.17 	& 2.64	& 3.15	& 4.02	& 2.65	& 3.48	& 2.41\footnotemark[7], 1.20\footnotemark[15]	\\
$\downarrow$-HOMO-1			& $^{3}\Sigma^{+}$	& 3.96\footnotemark[4]	& 2.71  	& 3.35 	& 3.47	& 3.31	& 3.00	& 4.11	& 4.51	& 4.15\footnotemark[7], 2.89\footnotemark[15]	\\
MAE							&						&		& 1.25	& 0.33	& 0.29	& 0.64	& 1.19	& 0.12	& 0.78	&	\\
\hline \hline
\end{tabular*}
\footnotetext[1]{Ref.~\onlinecite{Wu98}}
\footnotetext[2]{Ref.~\onlinecite{Wu97a}}
\footnotetext[3]{Ref.~\onlinecite{Wu97b}}
\footnotetext[4]{Ref.~\onlinecite{Moravec01}}
\footnotetext[5]{HOMO-1 and HOMO-2 are not included because our $GW$ calculations cannot account for two-electron transitions (see text).}
\footnotetext[6]{We chose this value from the Z band in the PES spectrum of CuO$^{-}$ (see text).}
\footnotetext[7]{Ref.~\onlinecite{Dai03} using 6-3111+G* basis sets}
\footnotetext[8]{Ref.~\onlinecite{Bridgeman00}}
\footnotetext[9]{Ref.~\onlinecite{Miliordos10} using the multi-reference configuration interaction (MRCI) method}
\footnotetext[10]{Ref.~\onlinecite{Gutsev00} using the B3LYP functional}
\footnotetext[11]{Ref.~\onlinecite{Gonzales00} using the B3LYP functional}
\footnotetext[12]{Ref.~\onlinecite{Bauschlicher98} using the B3LYP functional}
\footnotetext[13]{Ref.~\onlinecite{Korbel14} using the $G_{0}W_{0}@\text{PBE}$0 method}
\footnotetext[14]{Ref.~\onlinecite{Walsh99} using the B3LYP functional}
\footnotetext[15]{Ref.~\onlinecite{Hung17b} using the $G_{0}W_{0}@\text{PBE}$ method}
\footnotetext[16]{Ref.~\onlinecite{Xian00} using the CCSD(T) method}
\footnotetext[17]{Ref.~\onlinecite{Daoudi99}}
\footnotetext[18]{Ref.~\onlinecite{Bagus83} using the single and double excitation configuration interaction (SDCI) method}
\end{table*}


\section{Results and Discussion}
  
In this section, we compare our $GW$ calculations to anion PES experiments,~\cite{Wu98,Wu97a,Wu97b,Moravec01} focusing especially on the first IE, the lowest 3$d$-electron binding energy, and the orbital order. We present our results from two approaches seperately: First, we discuss non-self-consistent $GW$ with different starting-points (namely, $G_0W_0@\text{PBE}\alpha$ calculations as $\alpha$ is varied in steps of 0.25 from 0 to 1), and then, we discuss eigenvalue self-consistent $GW$ ($G_nW_0$ and $G_nW_n$) with PBE starting point. We only briefly discuss our $GW$ results for the starting-point--self-consistency hybrid approach, because (i) fundamentally, we found that the hybrid approach does not give any better results than the two separate approaches, and (ii) practically, the hybrid approach inherits disadvantages from both approaches.

ScO$^{-}$, TiO$^{-}$, CuO$^{-}$, and ZnO$^{-}$ are similar but different systems in several aspects. First, ScO$^{-}$ and CuO$^{-}$ are closed-shell systems, whereas TiO$^{-}$ and ZnO$^{-}$ are open-shell systems. Second, ScO$^{-}$ and TiO$^{-}$ have partially filled 3$d$ shells, while CuO$^{-}$ and ZnO$^{-}$ have completely filled 3$d$ shells. Third, TiO$^{-}$ has a shallow 3$d$ state, but CuO$^{-}$ and ZnO$^{-}$ have deep 3$d$ states. Fourth, 3$d$-electron photodetachment transitions are observed in TiO$^{-}$ and CuO$^{-}$, but not in ScO$^{-}$ and ZnO$^{-}$. Fifth, CuO$^{-}$ has strong 3$d$ character in HOMO, but ScO$^{-}$, TiO$^{-}$, and ZnO$^{-}$ have weak 3$d$ character in HOMO. Last, two-electron transitions are observed in ScO$^{-}$, but not in TiO$^{-}$, CuO$^{-}$, and ZnO$^{-}$. Due to these similarities and differences, TMO anions are an ideal set of systems for assessment of the performance of $GW$ schemes.

Table~\ref{tab:Orbital.Analysis} shows the amount of TM 3$d$ character in all molecular orbitals considered in this work, obtained from CN=2 gKS-PBE and gKS-PBE$\alpha$($\alpha$=1.00) using the Mulliken population analysis. We see that gKS-PBE $\uparrow$-HOMO of TiO$^{-}$ and gKS-PBE HOMO-2 of CuO$^{-}$ have entirely TM 3$d$ character. Fig.~\ref{Orbital.Plot} shows the contour plots of all molecular orbitals considered in this work, obtained from CN=2 gKS-PBE. It is clearly seen that $\uparrow$-HOMO of TiO$^{-}$ and HOMO-2 of CuO$^{-}$ are strongly localized on Ti and Cu, respectivley. Table~\ref{tab:BE.exp.ours.others} summarizes our $GW$ calculations with comparison to PES experiments and existing calculations in the literature.

Throughout this work, we use only gKS-PBE TM 3$d$ character except when we discuss the subtle competition between direct and indirect relativistic effects, because the EXX amount has a small effect on TM 3$d$ character. Also, throughout this work, we use only the gKS-PBE orbital order to avoid confusion. The orbital order depends strongly on the amount of EXX (e.g. PBE vs HF) and the level of theory (e.g. DFT vs $GW$).~\cite{vanSetten13,Govoni18} For example, gKS-PBE HOMO of CuO$^{-}$ corresponds to gKS-PBE$\alpha$($\alpha$=1.00) HOMO-1 and $G_{0}W_{0}@\text{PBE}\alpha$($\alpha$=1.00) HOMO of CuO$^{-}$, as shown in supplementary material and Fig.~\ref{G0W0.EXX.CuO.ZnO}. In this work, gKS-PBE and $G_{0}W_{0}@\text{PBE}$ were found to have the same orbital order.


\begin{figure*}
\begin{tabular}{c c}
{\includegraphics[trim=0mm 0mm 0mm 10mm, clip, width=0.40\textwidth]{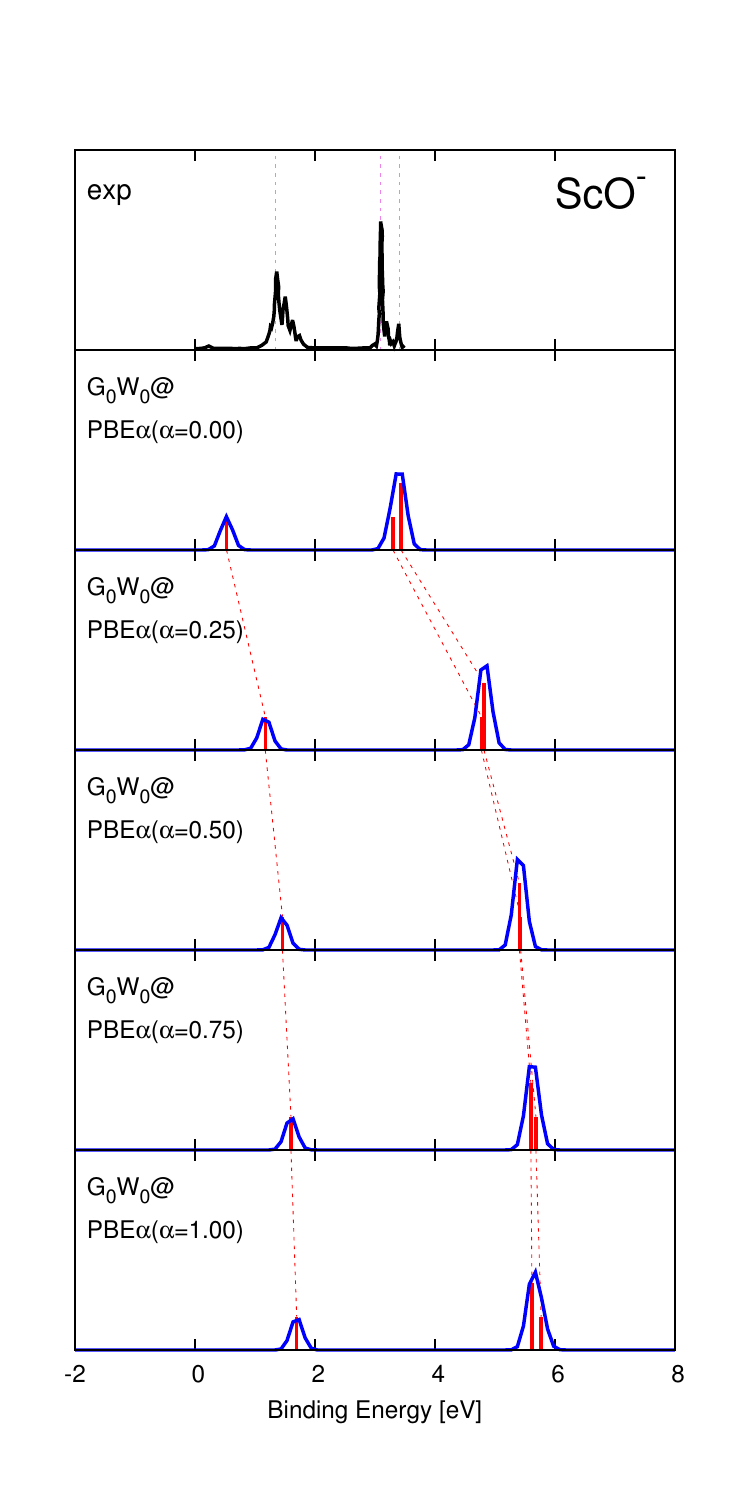}}
{\includegraphics[trim=0mm 0mm 0mm 10mm, clip, width=0.40\textwidth]{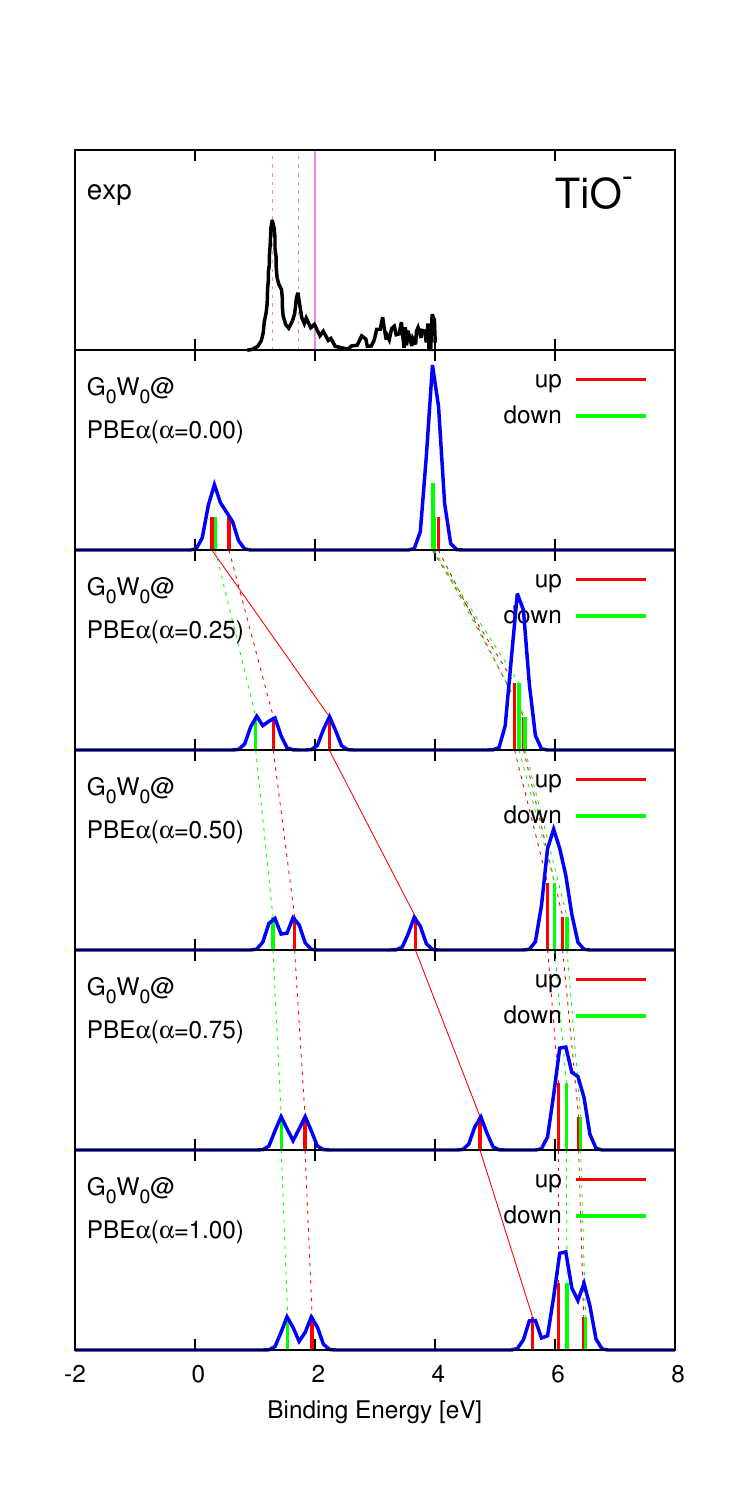}}
\end{tabular}
\caption{(Color online) Effect of the EXX amount in the $G_{0}W_{0}$ starting point on the electronic structure of ScO$^{-}$ and TiO$^{-}$. A Gaussian distribution function with a smearing width of 0.1~eV is used to broaden the spectra.}
\label{G0W0.EXX.ScO.TiO}
\end{figure*}

\begin{figure*}
\begin{tabular}{c c}
{\includegraphics[trim=0mm 0mm 0mm 10mm, clip, width=0.40\textwidth]{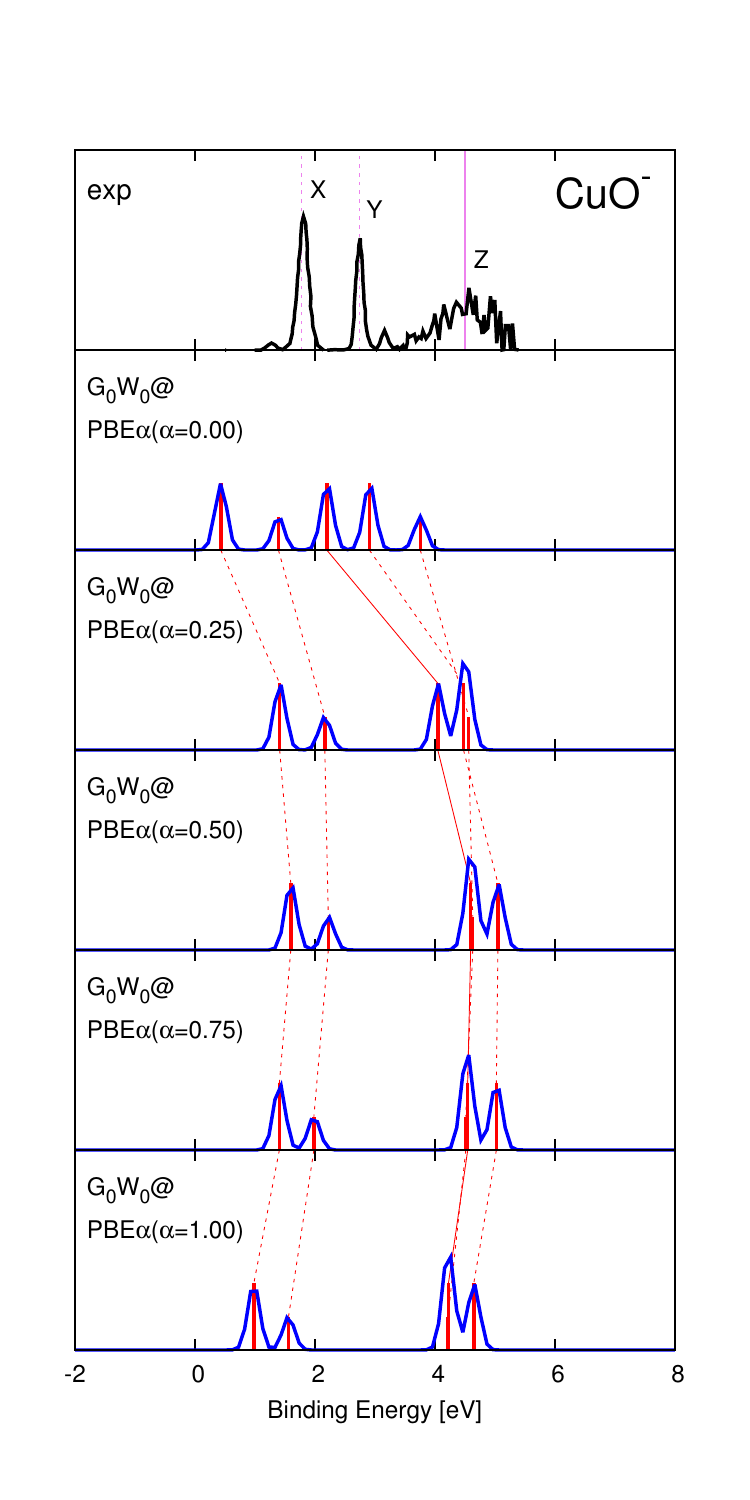}}
{\includegraphics[trim=0mm 0mm 0mm 10mm, clip, width=0.40\textwidth]{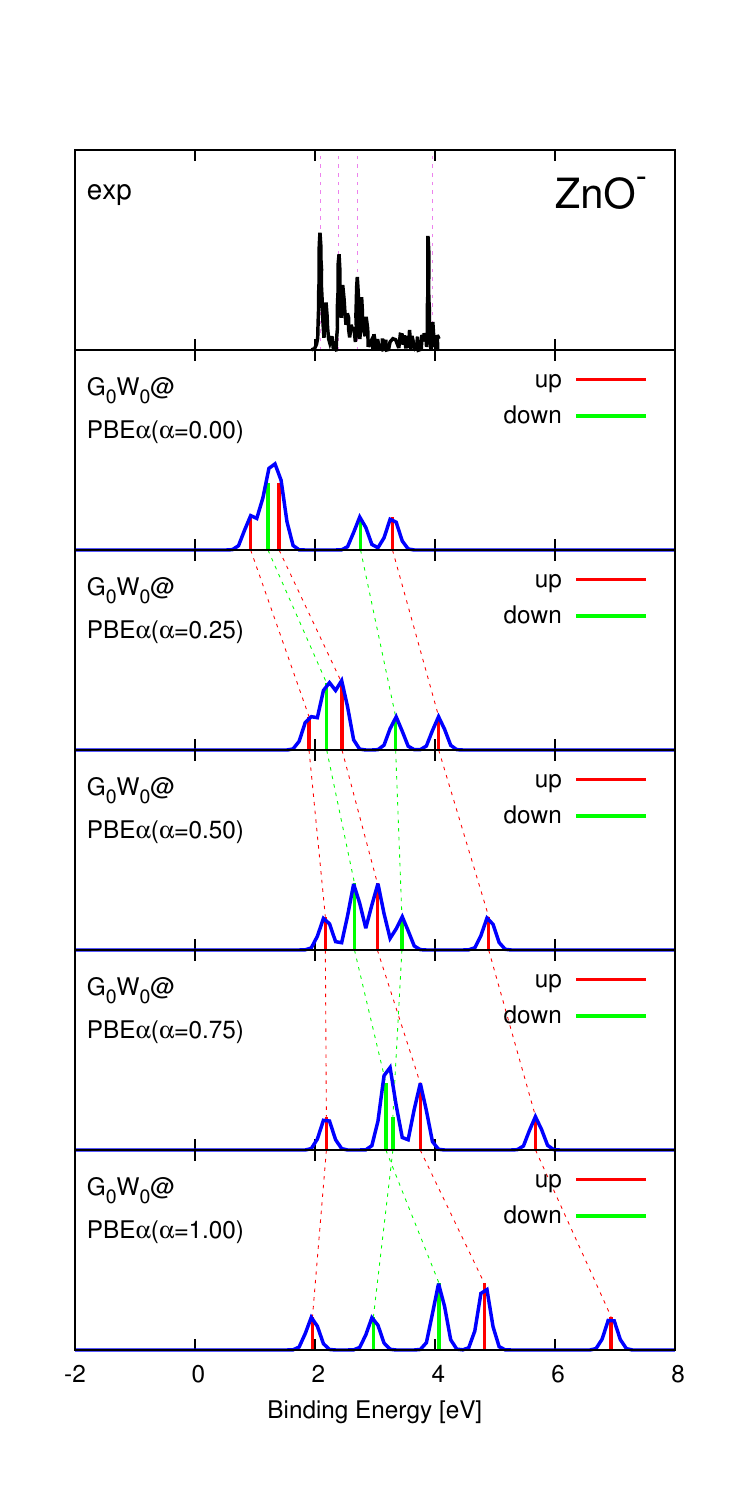}}
\end{tabular}
\caption{(Color online) Effect of the EXX amount in the $G_{0}W_{0}$ starting point on the electronic structure of CuO$^{-}$ and ZnO$^{-}$. A Gaussian distribution function with a smearing width of 0.1~eV is used to broaden the spectra.}
\label{G0W0.EXX.CuO.ZnO}
\end{figure*}

\subsection{$G_{0}W_{0}$ Starting Points} \label{sec:results.G0W0}

Figs.~\ref{G0W0.EXX.ScO.TiO} and \ref{G0W0.EXX.CuO.ZnO} show PES and $G_{0}W_{0}@\text{PBE}\alpha(0.00 \le \alpha \le 1.00)$ QP spectra of ScO$^{-}$, TiO$^{-}$, CuO$^{-}$, and ZnO$^{-}$. In PES spectra, vertial dashed and solid lines represent experimental $sp$- and $d$-electron binding energies, respectively. In $GW$ spectra, oblique dashed and solid lines track calculated $sp$- and $d$-electron binding energies, respectively. In Figs.~\ref{G0W0.EXX.ScO.TiO} and \ref{G0W0.EXX.CuO.ZnO}, we find a few general trends common in all TMO anions considered in this work: (i) no $G_{0}W_{0}@\text{PBE}\alpha(0.00 \le \alpha \le 1.00)$ results are in perfect agreement with experiment, (ii) $G_{0}W_{0}@\text{PBE}$ underestimates the IE of TMO anions by $\sim$1~eV, which is larger than the typical underestimation for $sp$ molecules ($\sim$0.5~eV),~\cite{vanSetten15,Maggio17,Govoni18} and (iii) $G_{0}W_{0}@\text{PBE}\alpha(0.25 \le \alpha \le 0.50)$ reduces it to $\sim$0.1~eV. In the following, we analyze each TMO anion individually.

\subsubsection{ScO$^{-}$}

Scandium is the first transition metal and has only one 3$d$ electron. DFT and CCSD(T) calculations in Refs.~\onlinecite{Gonzales00,Gutsev00} confirmed the ground state of ScO$^{-}$ as $^{1}\Sigma^{+}$ ($8\sigma^{2}3\pi^{4}9\sigma^{2}$), correcting the wrongly assumed state $^{3}\Delta^{-}$ ($8\sigma^{2}3\pi^{4}9\sigma^{1}1\delta$) in Ref.~\onlinecite{Wu98}. There is no 3$d$ peak or band in the PES spectrum of ScO$^{-}$, and the top three valence molecular orbitals have weak Sc 3$d$ character (6\%, 20\%, and 18\%, respectively), as shown in Table~\ref{tab:Orbital.Analysis}.

In the left panel of Fig.~\ref{G0W0.EXX.ScO.TiO}, we see that for HOMO-1 and HOMO-2 of ScO$^{-}$, $G_{0}W_{0}@\text{PBE}$ binding energies slightly overestimate PES ones (by 0.20 and 0.02~eV for HOMO-1 and HOMO-2, respectively), whereas $G_{0}W_{0}@\text{PBE}\alpha(0.25 \le \alpha \le 1.00)$ binding energies significantly overestimate PES ones by $\sim$2~eV [e.g. for HOMO-1, $G_{0}W_{0}@\text{PBE}\alpha$($\alpha$=0.25) and $G_{0}W_{0}@\text{PBE}\alpha$($\alpha$=1.00) binding energies are greater than PES ones by 1.67 and 2.72~eV, respectively]. This seems to suggest that for HOMO-1 and HOMO-2 of ScO$^{-}$, $G_{0}W_{0}@\text{PBE}$ performs better than $G_{0}W_{0}@\text{PBE}\alpha(0.25 \le \alpha \le 1.00)$, but this is not the case due to the nature of the corresponding peaks in the PES experiment. Ref.~\onlinecite{Gonzales00} suggests that the second and third peaks in the PES epectrum of ScO$^{-}$ are likely due to two-electron transitions from $8\sigma^{2}3\pi^{4}9\sigma^{2}$ ($^{1}\Sigma^{+}$ ScO$^{-}$) to $8\sigma^{2}3\pi^{4}10\sigma$ ($B^{2}\Sigma^{+}$ ScO) and to $8\sigma^{2}3\pi^{4}1\delta$ ($A'^{2}\Delta$ ScO) states, respectively, which $GW$ calculations for quasiparticle excitations cannot account for. In other words, the seemingly excellent agreement between $G_{0}W_{0}@\text{PBE}$ and PES binding energies for HOMO-1 and HOMO-2 of ScO$^{-}$ is accidental. Therefore, we exclude HOMO-1 and HOMO-2 of ScO$^{-}$ from our evaluation of the performance of $GW$ schemes in the following.

We also see that as $\alpha$ increases, $G_{0}W_{0}@\text{PBE}\alpha$ IE always increases, but this happens at different rates: As $\alpha$ increases from 0.00 to 0.25, it increases rapidly, whereas as $\alpha$ increases from 0.25 to 1.00, it increases slowly. The weak sensitivity of $G_{0}W_{0}@\text{PBE}\alpha(0.25 \le \alpha \le 1.00)$ IE to a change in $\alpha$ gives a large margin for an optimal amount of EXX: 25\%--100\%.


\subsubsection{TiO$^{-}$}

Titanium is the second transition metal and has two 3$d$ electrons. Several theoretical studies in Refs.~\onlinecite{Walsh99,Gutsev00,Gutsev03,Dai03,Miliordos10} confirmed $9\sigma^{2}\delta^{1}$ ($^{2}\Delta$) as the ground-state electron configuration of TiO$^{-}$, correcting the wrongly assigned configuration $9\sigma^{1}\delta^{2}$ ($^{1}\Sigma^{+}$) in Ref.~\onlinecite{Wu97a}. Unlike ScO$^{-}$, which has an empty $\delta$ shell, TiO$^{-}$ has one 3$d$ electron in the $\delta$ shell. The transition of the 3$d$ electron from $9\sigma^{2}\delta^{1}$ ($^{2}\Delta$ TiO$^{-}$) to $9\sigma^{2}$ ($^{1}\Sigma^{+}$ TiO) states produces the third peak in the PES spectrum of TiO$^{-}$ at 2.0~eV. In the $G_{0}W_{0}@\text{PBE}\alpha$ QP spectrum of TiO$^{-}$, $\uparrow$-HOMO is of entirely Ti 3$d$ character, as shown in Table~\ref{tab:Orbital.Analysis}.

The right panel of Fig.~\ref{G0W0.EXX.ScO.TiO} clearly shows that $G_{0}W_{0}@\text{PBE}\alpha(0.00 \le \alpha \le 1.00)$ binding energy for $\uparrow$-HOMO of TiO$^{-}$ is much more sensitive to a change in $\alpha$ than those for other occupied molecular orbitals with mainly $sp$ character, as shown in Table~\ref{tab:Orbital.Analysis}. The orbital-character-dependent sensitivity of $G_{0}W_{0}@\text{PBE}\alpha$ binding energy to a change in $\alpha$ causes a couple of problems. First, $G_{0}W_{0}@\text{PBE}$ underestimates the IE and the 3$d$-electron binding energy of TiO$^{-}$ non-uniformly (by 0.99 and 1.74~eV, respectively), leading to the wrong orbital order. In other words, $G_{0}W_{0}$ does not correct the wrong orbital order produced by PBE. Second, the $G_{0}W_{0}$ starting-point approach does not give accurate results for both the IE and the 3$d$-electron binding energy of TiO$^{-}$ at the same time. For example, $G_{0}W_{0}@\text{PBE}\alpha$($\alpha$=0.50) gives a better result for the IE of TiO$^{-}$ by 0.29~eV, but a worse result for the 3$d$-electron binding energy of TiO$^{-}$ by 1.46~eV, than $G_{0}W_{0}@\text{PBE}\alpha$($\alpha$=0.25). This type of behavior is not uncommon in $GW$ predictions for transition metal oxides; for example, no existing $GW$ scheme can accurately reproduce both the bandgap and the $d$-band position in the band structure of bulk ZnO at the same time.~\cite{Shishkin07,Fuchs07,Klimes14}

The increase in $\alpha$ from 0 to 1 has a similar effect on $G_{0}W_{0}@\text{PBE}\alpha$ IE of both ScO$^{-}$ and TiO$^{-}$: For both ScO$^{-}$ and TiO$^{-}$, $G_{0}W_{0}@\text{PBE}\alpha(0.25 \le \alpha \le 1.00)$ reduces the underestimation of IE by $G_{0}W_{0}@\text{PBE}$ from $\sim$1~eV to $\sim$0.1~eV [e.g. $G_{0}W_{0}@\text{PBE}\alpha$($\alpha$=0.25) reduces the difference in IE between PES and $G_{0}W_{0}@\text{PBE}$ from 0.84~eV to 0.20~eV and from 0.99~eV to 0.30~eV, respectively]. However, unlike ScO$^{-}$, the strong sensitivity of $G_{0}W_{0}@\text{PBE}\alpha$ 3$d$-electron binding energy of TiO$^{-}$ to a change in $\alpha$ gives a small margin for an optimal amount of EXX: $\sim$25\%. 

\subsubsection{CuO$^{-}$}

Copper is the 11th transition metal and has ten 3$d$ electrons. DFT calculations in Ref.~\onlinecite{Gutsev00} confirmed the ground state of CuO$^{-}$ as $^{1}\Sigma^{+}$ with the electron configuration of $3d^{10}2p\sigma^{2}2p\pi^{4}$. There are three bands (named as X, Y, and Z in Ref.~\onlinecite{Wu97b}) in the PES spectrum of CuO$^{-}$, as shown in the top of the left panel of Fig.~\ref{G0W0.EXX.CuO.ZnO}. Ref.~\onlinecite{Wu97b} suggested that the photodetachment transition of 3$d$ electrons ($3d\delta^{4}3d\pi^{4}3d\sigma^{2}$) from $^{1}\Sigma^{+}$ CuO$^{-}$ $3d^{10}2p\sigma^{2}2p\pi^{4}$ to Z CuO $3d^{9}2p\sigma^{2}2p\pi^{4}$ states produces the broad Z band in the PES spectrum of CuO$^{-}$ at $\sim$4.5~eV (which we selected from the position of the highest peak in the Z band) and assumed that the Z band is unusually broad likely due to a large geometry change from the anion to the neutral. In the $G_{0}W_{0}@\text{PBE}\alpha$ QP spectrum of CuO$^{-}$, HOMO-2 is of entirely Cu 3$d$ character, as shown in Table~\ref{tab:Orbital.Analysis}.

In the left panel of Fig.~\ref{G0W0.EXX.CuO.ZnO}, we see that $G_{0}W_{0}@\text{PBE}\alpha(0.00 \le \alpha \le 0.50)$ binding energy for HOMO-2 of CuO$^{-}$ is more sensitive to a change in $\alpha$ than those for other occupied molecular orbitals with weaker Cu 3$d$ character than HOMO-2, as shown in Table~\ref{tab:Orbital.Analysis}, and $G_{0}W_{0}@\text{PBE}\alpha(0.50 \le \alpha \le 0.75)$ gives good results for the IE and the 3$d$-electron binding energy (corresponding to HOMO and HOMO-2, respectively) of CuO$^{-}$ at the same time. Scalar relativistic effects in ECP reduce $G_{0}W_{0}@\text{PBE}\alpha$($\alpha$=0.50) and $G_{0}W_{0}@\text{PBE}\alpha$($\alpha$=0.75) binding energies for HOMO-2 of CuO$^{-}$ by 0.31 and 0.24~eV, as shown in supplementary material, without changing the conclusion that $G_{0}W_{0}@\text{PBE}\alpha(0.50 \le \alpha \le 0.75)$ gives good results for the 3$d$-electron binding energy of CuO$^{-}$. Like TiO$^{-}$, the orbital-character-dependent sensitivity of $G_{0}W_{0}@\text{PBE}\alpha$ binding energy to a change in $\alpha$ causes $G_{0}W_{0}@\text{PBE}$ to underestimate the IE and the 3$d$-electron binding energy of CuO$^{-}$ non-uniformly (by 1.38 and 2.32~eV, respectively). $G_{0}W_{0}@\text{PBE}\alpha(0.50 \le \alpha \le 1.00)$ binding energies for all valence molecular orbitals considered in this work are weakly sensitive to a change in $\alpha$. This trend suggests that PBE$\alpha(0.50 \le \alpha \le 1.00)$ orbitals with large amounts of EXX are good for localized states of CuO$^{-}$ with strong 3$d$ character [i.e. for CuO$^{-}$, PBE$\alpha(0.50 \le \alpha \le 1.00)$ wavefunctions are close to QP ones], and is consistent with the relatively good performance of HF on molecules with weak screening.~\cite{Rostgaard10,Huser13}


\subsubsection{ZnO$^{-}$}

Zinc is the 12th transition metal and has ten 3$d$ electrons. Zinc is rather distinct from other first row transition metals due to its closed-shell electron configuration. In other words, zinc is more similar to alkaline earth metals than other transition metals because Zn 3$d$ electrons generally do not participate in bonding.~\cite{Fancher98} DFT calculations in Ref.~\onlinecite{Gutsev00} confirmed the ground-state electron configuration of ZnO$^{-}$ as $^{2}\Sigma^{+}$ $10\sigma^{1}9\sigma^{2}4\pi^{4}\delta^{4}$. There are four bands in the PES spectrum of ZnO$^{-}$, as shown in the top of the right panel of Fig.~\ref{G0W0.EXX.CuO.ZnO}. The photodetachment of 3$d$ electrons is not measured in the PES experiment due to insufficient photon energy of 4.66~eV.~\cite{Moravec01} Unlike CuO$^{-}$, all valence molecular orbitals in the $G_{0}W_{0}@\text{PBE}\alpha$ QP spectrum of ZnO$^{-}$ have weak Zn 3$d$ character, as shown in Table~\ref{tab:Orbital.Analysis}.


The right panel of Fig.~\ref{G0W0.EXX.CuO.ZnO} shows that unlike CuO$^{-}$, $G_{0}W_{0}@\text{PBE}$ underestimates electron binding energies for all valence molecular orbitals of ZnO$^{-}$ uniformly (e.g. by 1.18, 1.35, 1.23, and 1.25~eV for $\uparrow$-HOMO, $\uparrow$-HOMO-1, $\downarrow$-HOMO, and $\downarrow$-HOMO-1, respectively) possibly because all valence molecular orbitals have similar Zn 3$d$ character and thus their $G_{0}W_{0}@\text{PBE}\alpha$ binding energies have similar sensitivity to a change in $\alpha$. $G_{0}W_{0}@\text{PBE}\alpha$($\alpha$=0.50) gives good results for the IE and the orbital order of ZnO$^{-}$ at the same time. Unlike CuO$^{-}$, $G_{0}W_{0}@\text{PBE}\alpha(0.50 \le \alpha \le 1.00)$ binding energies for all valence molecular orbitals, except for $\uparrow$-HOMO and $\downarrow$-HOMO-1, are strongly sensitive to a change in $\alpha$. This trend can be explained in terms of the importance of spin-splitting in open-shell molecules (as has been discussed in Ref.~\onlinecite{Shi18} for CuO$_{2}^{-}$ molecule).


\begin{figure*}
\begin{tabular}{c c}
{\includegraphics[trim=0mm 0mm 0mm 10mm, clip, width=0.40\textwidth]{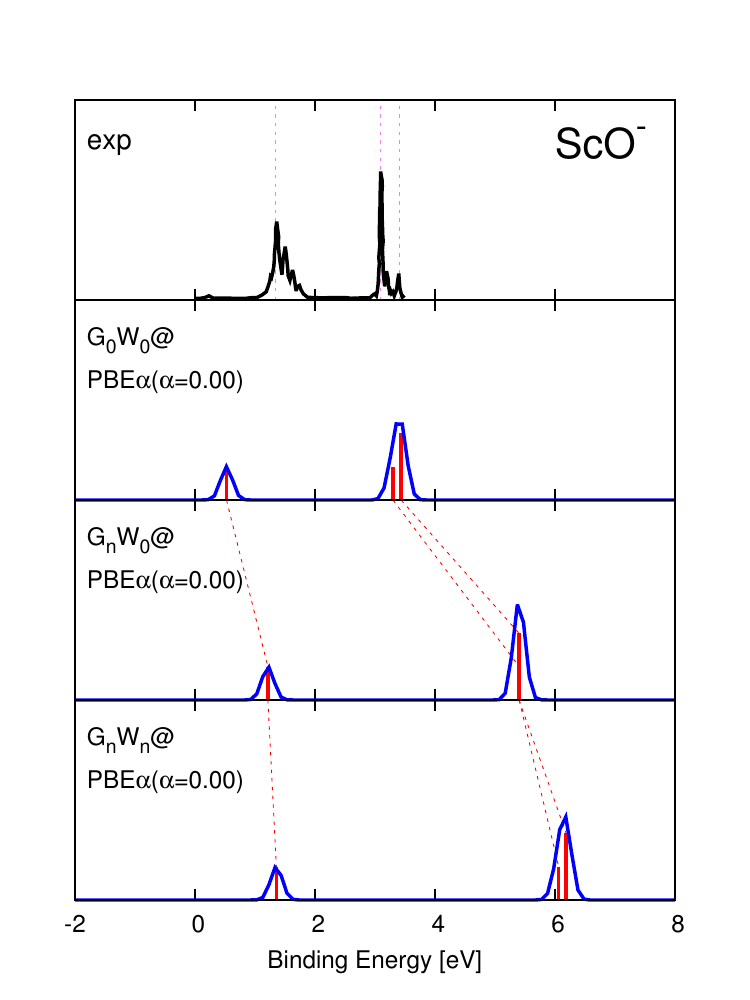}}
{\includegraphics[trim=0mm 0mm 0mm 10mm, clip, width=0.40\textwidth]{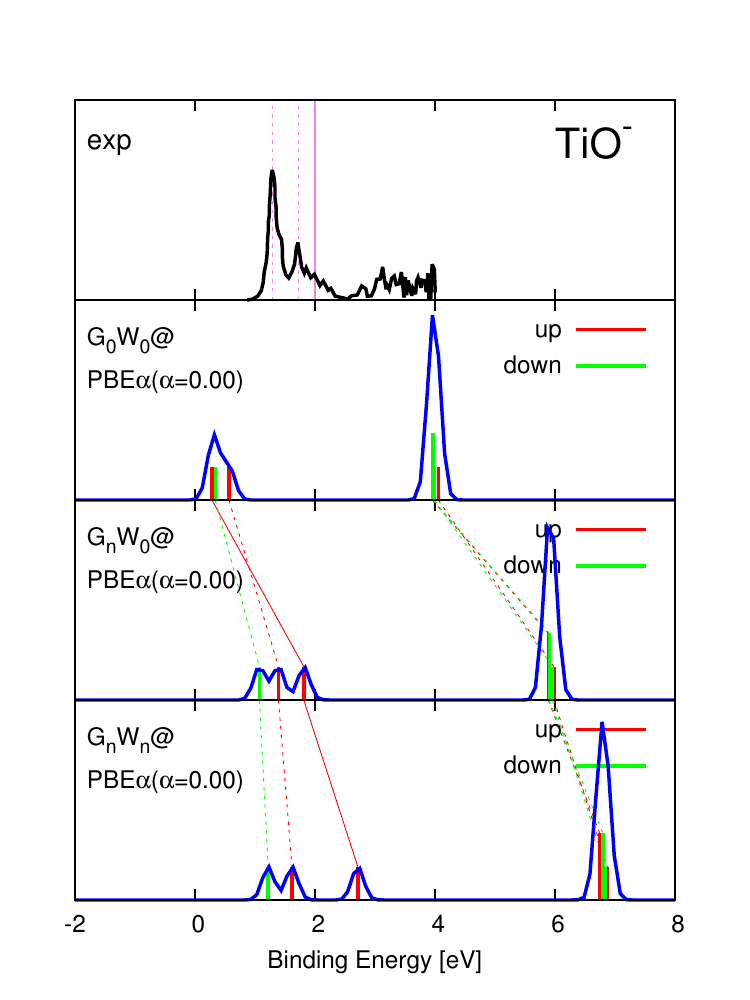}} \\
{\includegraphics[trim=0mm 0mm 0mm 10mm, clip, width=0.40\textwidth]{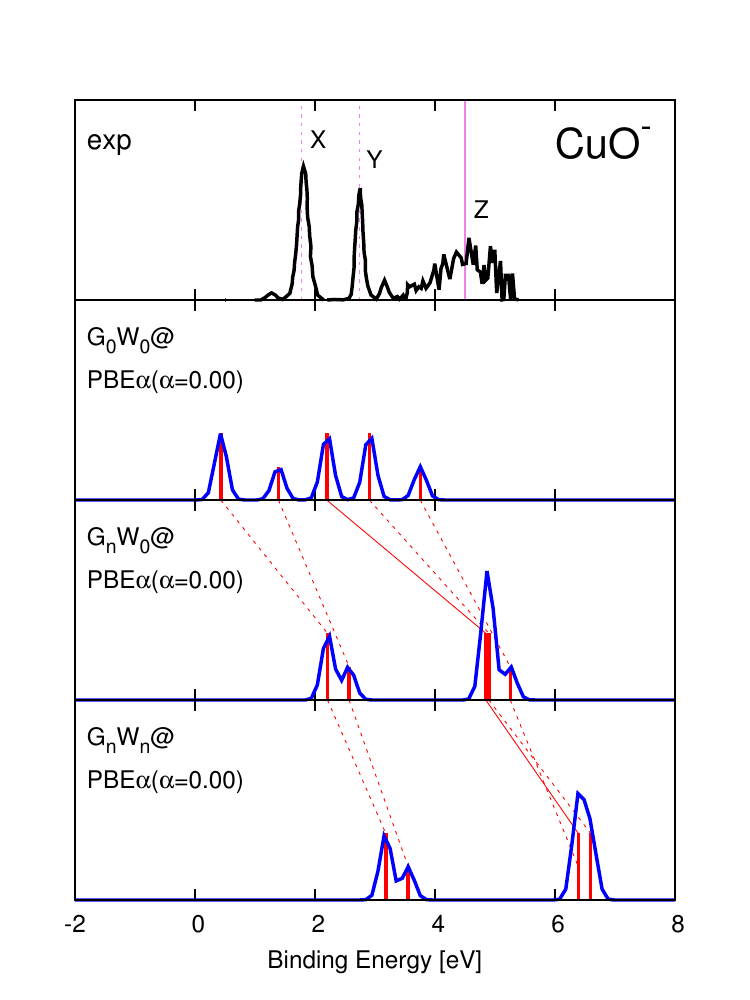}}
{\includegraphics[trim=0mm 0mm 0mm 10mm, clip, width=0.40\textwidth]{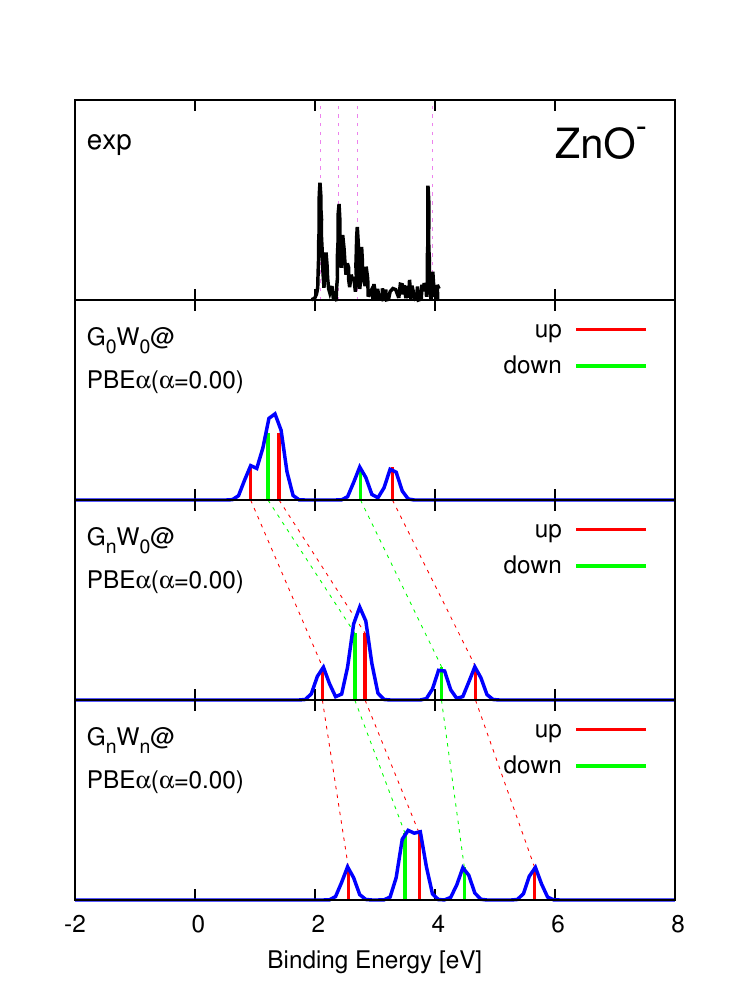}}
\end{tabular}
\caption{(Color online) Effect of the ev$GW$ self-consistency level on the electronic structure of ScO$^{-}$, TiO$^{-}$, CuO$^{-}$, and ZnO$^{-}$. A Gaussian distribution function with a smearing width of 0.1~eV is used to broaden the spectra.}
\label{GW.EXX000}
\end{figure*}

\subsection{ev$GW$ Self-Consistency Levels} \label{sec:results.evGW}

Fig.~\ref{GW.EXX000} shows PES and ev$GW$ QP spectra of ScO$^{-}$, TiO$^{-}$, CuO$^{-}$, and ZnO$^{-}$. In the following, we analyze $G_{n}W_{0}@\text{PBE}$ and $G_{n}W_{n}@\text{PBE}$ results individually.

In Fig.~\ref{GW.EXX000}, we see that as the ev$GW$ self-consistency level increases from $G_{0}W_{0}$ to $G_{n}W_{n}$, $GW$ binding energies always increase, but this occurs at different rates: As the ev$GW$ self-consistency level increases from $G_{0}W_{0}$ to $G_{n}W_{0}$, $GW$ binding energies increase rapidly (e.g. the IE increases by 0.70, 0.75, 1.79, and 1.20~eV for ScO$^{-}$, TiO$^{-}$, CuO$^{-}$, and ZnO$^{-}$, respectivley), while as it increases from $G_{n}W_{0}$ to $G_{n}W_{n}$, they increase slowly (e.g. the IE increases by 0.14, 0.15, and 0.46~eV for ScO$^{-}$, TiO$^{-}$, and ZnO$^{-}$, respectivley) except for CuO$^{-}$ (0.98~eV), which will be discussed later. $G_{0}W_{0}@\text{PBE}$ always underestimates electron binding energies, whereas $G_{n}W_{n}@\text{PBE}$ generally overestimates them. $G_{n}W_{0}@\text{PBE}$ binding energies are always in between $G_{0}W_{0}@\text{PBE}$ and $G_{n}W_{n}@\text{PBE}$ ones and generally close to experiment. In other words, $G_{0}W_{0}@\text{PBE}$ and $G_{n}W_{n}@\text{PBE}$ act as lower and upper bounds for $G_{n}W_{0}@\text{PBE}$, generally producing over- and under-screenings, respectively. This trend of the ev$GW$ self-consistency approach in electronic structure of molecules is also observed in band structure of solids.~\cite{Shishkin07}


We also see that the ev$GW$ self-consistency has a strong effect on $GW$ binding energies for molecular orbitals with strong 3$d$ character (e.g. $\uparrow$-HOMO of TiO$^{-}$ and HOMO-2 of CuO$^{-}$).  For example, $G_{n}W_{0}@\text{PBE}$ reduces the underestimation errors of $G_{0}W_{0}@\text{PBE}$ in the IE and the 3$d$-electron binding energy of TiO$^{-}$ with respect to experiment from 0.99 and 1.74~eV to 0.24 and 0.17~eV, respectively. As a result, $G_{n}W_{0}@\text{PBE}$ corrects the wrong $G_{0}W_{0}@\text{PBE}$ orbital order in TiO$^{-}$. Another example is that $G_{n}W_{0}@\text{PBE}$ gives small ($\sim$0.1~eV) errors in electron binding energies for all valence molecular orbitals of ZnO$^{-}$, which are uniformly underestimated by $G_{0}W_{0}@\text{PBE}$ by $\sim$1~eV due to similarly weak Zn 3$d$ character. For ZnO$^{-}$, $G_{0}W_{0}@\text{PBE}$ and $G_{n}W_{0}@\text{PBE}$ yield mean absolute errors (MAEs) of 1.25 and 0.12~eV, respectively, as shown in Table~\ref{tab:BE.exp.ours.others}.

CuO$^{-}$ exhibits particularly large differences between $G_{n}W_{0}@\text{PBE}$ and $G_{n}W_{n}@\text{PBE}$ binding energies compared to other TMO anions. This trend is not associated with scalar relativistic effects in ECP, which reduce $G_{n}W_{0}@\text{PBE}$ and $G_{n}W_{n}@\text{PBE}$ binding energies by similar amounts (e.g. by 0.57 and 0.66~eV, respectively, for HOMO-2 of CuO$^{-}$, as shown in supplementary material). We attribute this trend to strong 3$d$ character in molecular orbitals of CuO$^{-}$. For example, CuO$^{-}$ has a larger difference between $G_{n}W_{0}@\text{PBE}$ and $G_{n}W_{n}@\text{PBE}$ IEs than ScO$^{-}$ (0.98 and 0.14~eV, respectively) possibly because CuO$^{-}$ has stronger 3$d$ character in HOMO than ScO$^{-}$ (23\% and 6\%, respectively, as shown in Table~\ref{tab:Orbital.Analysis}).


\subsection{Comparison of $G_{0}W_{0}$ starting-point and ev$GW$ self-consistency approaches}

From our results presented so far, it appears that both $G_{0}W_{0}$ starting-point and ev$GW$ self-consistent approaches can, in principle, be good $GW$ methods for finite systems: both $G_{0}W_{0}@\text{PBE}\alpha(0.25 \le \alpha \le 0.50)$ and $G_{n}W_{0}@\text{PBE}$ can reduce the large and orbital-character-dependent non-uniform errors for electron binding energies of TMO anions produced by $G_{0}W_{0}@\text{PBE}$ with respect to experiment from $\sim$1--2~eV to $\sim$0.1--0.5~eV. Ref.~\onlinecite{Fuchs07} obtained similar results for extended systems: both $G_{0}W_{0}@\text{PBE}\alpha$($\alpha$=0.25) and $G_{n}W_{0}@\text{PBE}$ give satisfactory results for the bandgap and the $d$-electron binding energy of solids, and drew the conclusions that (i) for accuracy, one can choose either $G_{0}W_{0}@\text{PBE}\alpha$($\alpha$=0.25) or $G_{n}W_{0}@\text{PBE}$ because they give similar results, but (ii) for efficiency, one may want to choose $G_{0}W_{0}@\text{PBE}\alpha$($\alpha$=0.25) over $G_{n}W_{0}@\text{PBE}$ because the former is computationally cheaper than the latter. However, in the case of molecular systems, we argue that $G_{n}W_{0}@\text{PBE}$ has several practical advantages over $G_{0}W_{0}@\text{PBE}\alpha$.

First, $G_{n}W_{0}@\text{PBE}$ does not contain system-dependent adjustable parameters. Unlike extended systems, there is no unique amount of EXX for the $G_{0}W_{0}$ starting point, which works well for all finite systems. For example, we showed in Section~\ref{sec:results.G0W0} that 25\% EXX is optimal for ScO$^{-}$ and TiO$^{-}$, whereas 50\% EXX is optimal for CuO$^{-}$ and ZnO$^{-}$. Also, it appears that atoms and small molecules require more amount of EXX than clusters and large molecules.~\cite{Blase16} Second, $G_{n}W_{0}@\text{PBE}$ is transferable between finite and extended systems. $G_{n}W_{0}@\text{PBE}$ works well for both molecules and solids (e.g. ZnO$^{-}$ anion and bulk ZnO, respectively).~\cite{Shishkin07} This greatly extends the range of applicability of the $GW$ method. For example, $G_{n}W_{0}@\text{PBE}$ may be applicable to solid-molecule hybrid systems such as molecular junctions and molecules adsorbed on solid surfaces.~\cite{Strange11} Also, $G_{n}W_{0}@\text{PBE}$ may be used for the study of quantum size effects in clusters because it is independent of the cluster size. Third, $G_{n}W_{0}@\text{PBE}$ is easy to use and reliable. Unlike PBE$\alpha(0.00 < \alpha \le 1.00)$, PBE is safe from the SCF convergence issue, and unlike $G_{0}W_{0}$, ev$GW$ with $Z = 1$ is immune to the $GW$ multi-solution issue. Therefore, $G_{n}W_{0}@\text{PBE}$ does not need manual, time-consuming, and error-prone tests to address the two issues, which are explained in detail in Section~\ref{sec:comp.details.test.results}.




Furthermore, $G_{n}W_{0}@\text{PBE}$ has a few desirable properties. One of them comes from the PBE part. PBE causes the smallest incomplete basis set error, as shown in Fig.~\ref{fit.NBF.HOMO.0.EXX000.EXX100}, allowing one to use smaller basis sets for the CBS limit, which makes $G_{n}W_{0}@\text{PBE}$ cheaper. Two desirable properties come from the $GW$ part. $G_{n}W_{0}$ (as well as $G_{n}W_{n}$) gives faster and more stable $GW$ convergence than QS$GW$ and depends more weakly on the choice of $\eta$ (e.g. we used a single value of $\eta$ for ev$GW$ in this work), as discussed in Section~\ref{sec:test.evGW.QP.energy}. Also, $G_{n}W_{0}$ is cheaper than $G_{n}W_{n}$, as pointed out in Ref.~\onlinecite{Shishkin07} and discussed in Section~\ref{sec:background.evGW.QP.energy}. In fact, $G_{n}W_{0}$ is the cheapest self-consistent $GW$ scheme.



One may argue that $G_{0}W_{0}@\text{PBE}\alpha$ should be a choice of $GW$ methods because it is computationally more efficient than $G_{n}W_{0}@\text{PBE}$ by the number of self-consistent $G_{n}W_{0}$ iterations. However, as discussed in Section~\ref{sec:background.evGW.QP.energy} and supplementary material, this is not the case since the compute time difference between $G_{0}W_{0}@\text{PBE}\alpha$ and $G_{n}W_{0}@\text{PBE}$ does not depend only on the number of $G_{n}W_{0}$ iterations; there are other factors such as the number of eigenvalues to update for $\epsilon^{\text{G}_{n}\text{W}_{0}}_{m}$, the number of frequency points to use for $\Sigma_{\text{c}}(\omega)$, the number of $\Delta\omega$ and $\eta$ values to test for $\epsilon^{\text{G}_{0}\text{W}_{0}}_{m}$, and the number of initial guess wavefunctions to test for gKS calculations. Some factors can cancel each other out; for example, $G_{n}W_{0}@\text{PBE}$ requires a few $G_{n}W_{0}$ iterations, but one typically needs to test a few $\eta$ values for $G_{0}W_{0}@\text{PBE}\alpha$. In other words, when all factors are taken into account, the total compute time to obtain reliable and reproducible QP spectra at $G_{0}W_{0}@\text{PBE}\alpha$ and $G_{n}W_{0}@\text{PBE}$ levels of theory can be comparable, as is especially the case for open-shell systems.


\begin{table*}
\caption{Optimal amount of EXX in the $G_{0}W_{0}$ starting point for gas-phase small molecules (highlighted in bold).}
\label{tab:G0W0.EXX.opt}
\begin{tabular*}{1.00\textwidth}{ @{\extracolsep{\fill}} l | c c c c c }
\hline \hline
Reference			& K\"{o}rbel \textit{et al.}~\cite{Korbel14} & Bruneval \textit{et al.}~\cite{Bruneval13} & Kaplan \textit{et al.}~\cite{Kaplan16} & Rostgaard \textit{et al.}~\cite{Rostgaard10}	& This work	\\ \hline
Code				& FIESTA 				& MOLGW 					& TURBOMOLE			& GPAW					& MOLGW							\\
Optimal EXX			& \textbf{25\%} 			& \textbf{50\%} 				& \textbf{75\%}			& \textbf{100\%}			& \textbf{25--50\%}					\\
Tested EXX			& 25 \& 100\% 			& 0, 20, 25 \& 50\% 			& 0, 25 \& 75\%			& 0 \& 100\%			& 0, 25, 50, 75 \& 100\%				\\
System 				& 39 closed-shell 3,4,5$d$ & 34 closed-shell $sp$\footnotemark[1] & 29 closed-shell $sp$ & 34 closed-shell $sp$\footnotemark[1] & 4 closed- \& open-shell 3$d$ \\
           				& \& 9 closed-shell $sp$	&							&						&						&                                            		\\
System size			& 2--7 atoms 			& 2--8 atoms					& 2--18 atoms			& 2--8 atoms				& 2 atoms							\\
Property				& HOMO \& LUMO		& HOMO						& HOMO \&				& HOMO					& HOMO-$n$ ($n$ = 0, 1, ...)			\\
					&						&							& HOMO-$n$ ($n$ = 0, 1, ...)\footnotemark[2] & 	& (focusing on 3$d$ MO) 				\\
Reference data		& Experiment			& $\Delta$SCF\footnotemark[3]	& QS$GW$\footnotemark[4] & Experiment		& Experiment						\\
$\omega$ integration	& Contour deformation\footnotemark[5] & Fully analytic	& Fully analytic			& Fully analytic		 	& Fully analytic						\\
QP equation			& Linearization 			& Linearization 				& Spectral function 		& Linearization\footnotemark[6] \& 		& Graphical solution \&	\\
					&                    			&							&						& Spectral function\footnotemark[7]	& Spectral function 		\\
$\eta$				& Not available 			& Not available				& 0.001~eV				& 0~eV					& 0.002 or 0.005~Ha					\\
CBS limit			& Not used				& Not used					& Not used 				& Not used				& Used [employing Eq.~(\ref{eq:fit.NBF})]		\\
					& (CN=4 only)			& (CN=4 only)				& (CN=3 only)			& (CN=2 only)			& (CN=2,3,4,5)						\\
Potential				& ECP					& AE						& AE					& PAW\footnotemark[8]	& AE								\\
RI					& Used					& Not used					& Used					& Not applicable\footnotemark[9]		& Not used				\\       
\hline \hline
\end{tabular*}
\footnotetext[1]{The same set of molecules is used.}
\footnotetext[2]{For naphthalene only}
\footnotetext[3]{Ref.~\onlinecite{Bruneval13} showed that $\Delta$SCF using CCSD(T) with CN=4 causes an error of $\sim$0.1~eV in the IE of small $sp$ molecules with respect to experiment (the largest being 0.67~eV for NaCl).}
\footnotetext[4]{Ref.~\onlinecite{Bruneval12} showed that QS$GW$ with CN=5 causes a mean absolute error of 0.18~eV in the IE of the first row atoms with respect to experiment (the largest being $\sim$0.4~eV for O).}
\footnotetext[5]{Refs.~\onlinecite{vanSetten15,Golze18} showed that the contour deformation technique produces almost the same $GW$ self-energy as the fully analytic method for frontier and non-frontier orbitals, respectively.}
\footnotetext[6]{For 0\% EXX}
\footnotetext[7]{For 100\% EXX}
\footnotetext[8]{Projector-Augmented Wave}
\footnotetext[9]{GPAW uses augmented Wannier basis sets, whereas FIESTA, MOLGW, and TURBOMOLE use Gaussian basis sets.}
\end{table*}

\subsection{Comparison with results in the literature}

Some of our results for the performance of $G_{0}W_{0}$ starting-point and ev$GW$ self-consistency approaches in this work may seem to be at odds with some of the results in the literature. In this section, we discuss the origin of the apparent differences between them.

We begin with the $G_{0}W_{0}$ starting-point approach. Table~\ref{tab:G0W0.EXX.opt} summarizes a few selected results for the optimal amount of EXX in the $G_{0}W_{0}$ starting point out of numerous results, such as Refs.~\onlinecite{Marom12,Caruso16}, in the literature. Interestingly, we see that there is a wide range of EXX amounts from 25\% to 100\%, and Refs.~\onlinecite{Bruneval13,Rostgaard10} obtained different results (50\% and 100\%, respectively) from the same set of molecules. It seems that 75\% and 100\% are too large compared to our results: 25--50\%. One may guess that the large difference is due to implementation differences such as basis type (e.g. Gaussian vs PW) and frequency integration type (e.g. analytical vs numerical). However, Refs.~\onlinecite{Maggio17,Govoni18} showed that such implementation differences have little effect on $G_{0}W_{0}$ IE ($\sim$0.06~eV). There are a couple of other factors that have a stronger effect on $G_{0}W_{0}$ results than implementation differences. One factor is the choice of system and property. As shown in Section~\ref{sec:results.G0W0}, $G_{0}W_{0}@\text{PBE}\alpha(0.25 \le \alpha \le 1.00)$ IEs of $sp$ systems are slightly different (by $\sim$0.1~eV). Most existing $G_{0}W_{0}$ studies used the IE of $sp$-bonded systems to determine the optimal amount of EXX in the $G_{0}W_{0}$ starting point. The other factor is that the choice of QP equation solver and CBS extrapolation method. As shown in Section~\ref{sec:CBS} and Section~\ref{sec:test.G0W0.QP.energy}, the linearization method and the CBS extrapolation method (e.g. whether to extrapolate or not and which fitting function and basis set to use for extrapolation) can cause a difference in $G_{0}W_{0}$ IE on the order of $\sim$0.1~eV. Overall, the combination of the two factors gives a large margin for the optimal amount of EXX in the $G_{0}W_{0}$ starting point, and thus is likely to produce the wide range of amounts that exist in the literature.


\begin{figure*}
\begin{tabular}{c c c}
{\includegraphics[trim=0mm 0mm 0mm 0mm, clip, width=0.33\textwidth]{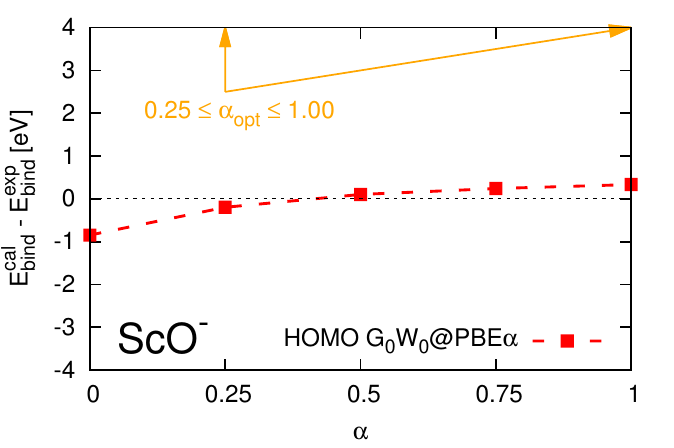}}
{\includegraphics[trim=0mm 0mm 0mm 0mm, clip, width=0.33\textwidth]{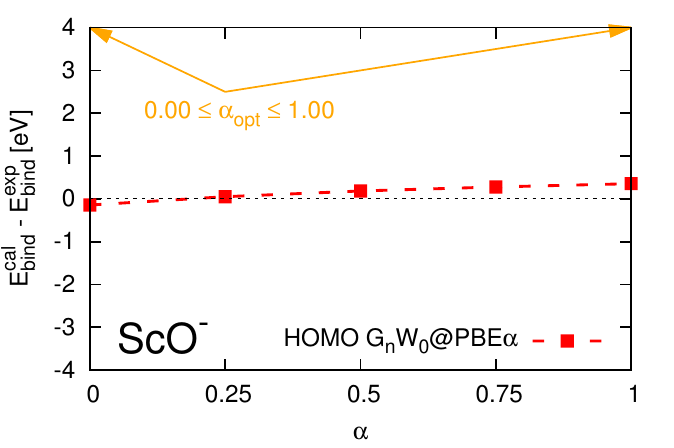}}
{\includegraphics[trim=0mm 0mm 0mm 0mm, clip, width=0.33\textwidth]{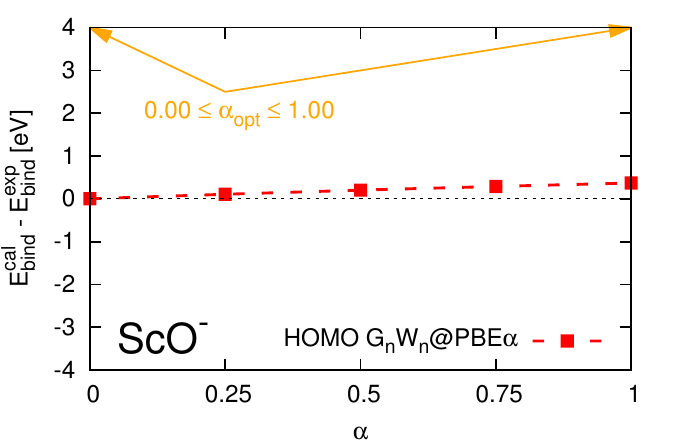}} \\
{\includegraphics[trim=0mm 0mm 0mm 0mm, clip, width=0.33\textwidth]{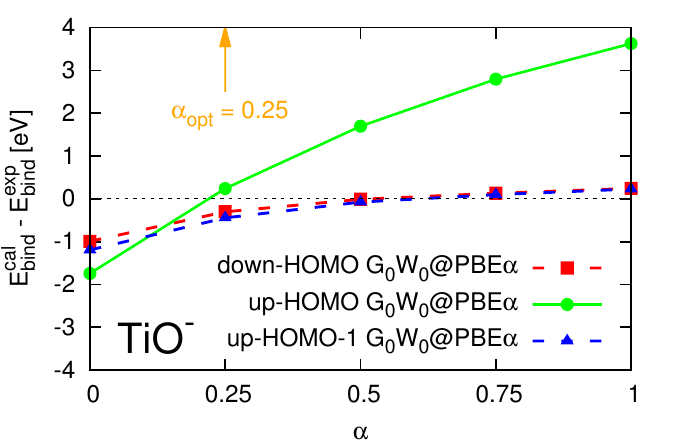}}
{\includegraphics[trim=0mm 0mm 0mm 0mm, clip, width=0.33\textwidth]{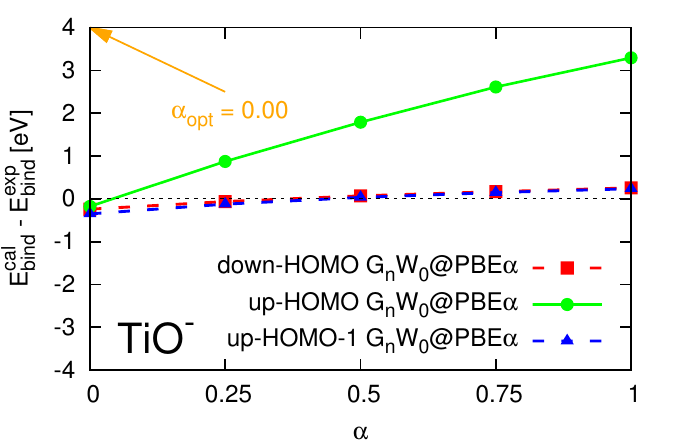}}
{\includegraphics[trim=0mm 0mm 0mm 0mm, clip, width=0.33\textwidth]{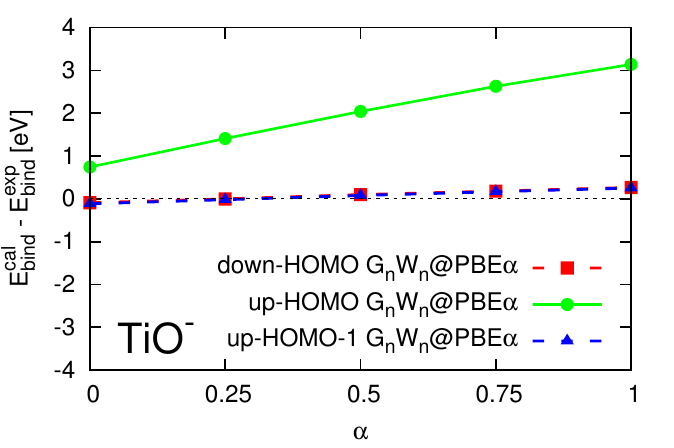}} \\
\end{tabular}
\caption{(Color online) Effect of the $GW$ starting point and the ev$GW$ self-consistency level on the electronic structure of ScO$^{-}$ and TiO$^{-}$. $E_{\text{bind}}^{\text{exp}}$ and $E_{\text{bind}}^{\text{cal}}$ represent experimental and calculated electron binding energies, respectively. Dashed and solid lines track $sp$- and $d$-electron binding energies, respectively. $\alpha_{\text{opt}}$ represents an optimal fraction of EXX in the $GW$ starting point.}
\label{GW.EXXopt}
\end{figure*}

Next, we move on to the ev$GW$ self-consistency approach, and discuss the origin of apparently conflicting ev$GW$ results for IE and starting-point dependency. First, Ref.~\onlinecite{Blase11} reported that the $G_{n}W_{n}$ approach with a local-density approximation starting point gives good results for the IE of large $sp$ molecules, whereas we found in Section~\ref{sec:results.evGW} that $G_{n}W_{0}@\text{PBE}$ gives satisfactory results for the electronic structure (including the IE) of small 3$d$ molecules. A comparison of ev$GW$ implementations in Ref.~\onlinecite{Blase11} and this work is provided in supplementary material. We believe that the main origin of the different results is the orbital-character-dependent sensitivity of ev$GW$ binding energy to a change in ev$GW$ self-consistency level. As shown in Section~\ref{sec:results.evGW}, $G_{n}W_{0}@\text{PBE}$ and $G_{n}W_{n}@\text{PBE}$ binding energies are slightly different for delocalized HOMO with weak 3$d$ character by $\sim$0.1~eV, but significantly different for localized HOMO with strong 3$d$ character by $\sim$1~eV. Unlike this work, Ref.~\onlinecite{Blase11} used the linearization method, employed pseudopotentials and RI, and did not use the CBS limit, but these cause small ($\sim$0.1~eV) differences in ev$GW$ IE, as shown in Section~\ref{sec:GW.test.results}. Accordingly, they are most likely not the reason for the large (0.98~eV) difference between $G_{n}W_{0}@\text{PBE}$ and $G_{n}W_{n}@\text{PBE}$ IEs of CuO$^{-}$. Second, Ref.~\onlinecite{Blase16} reported that in small water clusters, as the ev$GW$ self-consistency level increases, the ev$GW$ starting-point dependency decreases, whereas we found in Section~\ref{sec:test.evGW.QP.energy} that in TMO anions, $G_{n}W_{n}$ sometimes depends more strongly on the starting point than $G_{n}W_{0}$. As mentioned in Section~\ref{sec:test.evGW.QP.energy}, we believe that the orbital character influences the ev$GW$ starting-point dependency: for molecular orbitals with strong (weak) 3$d$ character, $G_{n}W_{n}$ depends more strongly (weakly) on the starting point than $G_{n}W_{0}$. Overall, without molecular orbitals with strong 3$d$ character (e.g. HOMO of CuO$^{-}$), our ev$GW$ results for IE and starting-point dependency in this work are consistent with those in Refs.~\onlinecite{Blase11,Blase16}.


To verify our idea about the origin of the seemingly different results between this work and the literature, we performed a simple test: (i) we chose ScO$^{-}$ and TiO$^{-}$ as our analogs of $sp$ molecules in the literature because their valence molecular orbitals have weak transition-metal character, except for $\uparrow$-HOMO of TiO$^{-}$ with entirely Ti 3$d$ character, (ii) we applied 15 different starting-point--self-consistency hybrid $GW$ schemes ($G_{0}W_{0}$, $G_{n}W_{0}$, and $G_{n}W_{n}$; 0\%, 25\%, 50\%, 75\%, and 100\% EXX) to them, and (iii) we searched for $GW$ schemes that give a reasonably small error of less than 0.5~eV in the IE and the 3$d$-electron binding energy with respect to experiment. Fig.~\ref{GW.EXXopt} shows the results of the test. We see that $GW$ IEs of ScO$^{-}$ and TiO$^{-}$ (red dashed lines) depend weakly on the starting point and the self-consistency level, giving a large margin for the choice of $GW$ schemes. 14 $GW$ schemes out of 15 ($G_{0}W_{0}@\text{PBE}$ is an exception as expected) give a small error (less than 0.5~eV), which explains why there are a large number of different good $GW$ schemes for the IE of $sp$ molecules in the literature. We also see that the $GW$ 3$d$-electron binding energy of TiO$^{-}$ (green solid lines) depends strongly on the starting point and the self-consistency level, yielding a small margin for the choice of $GW$ schemes. Only two $GW$ schemes ($G_{0}W_{0}@\text{PBE}$0 and $G_{n}W_{0}@\text{PBE}$) out of 15 give a small error (less than 0.5~eV), which is why we obtained a small number of good $GW$ schemes for the electronic structure of $d$ molecules in this work. Overall, we confirm that evaluation results for the performance of $GW$ schemes depend strongly on the choice of system and property (e.g. the IE with mainly $sp$ character vs the electronic structure containing $d$ states). 

\section{Summary and Conclusions}

In summary, we calculated the electronic structure of closed- and open-shell molecular anions with partially and completely filled 3$d$ shells (shallow and deep 3$d$ states, respectively) using various $GW$ schemes and compared calculated $GW$ QP spectra to anion PES experiments to evaluate the performance of the $GW$ approximation on both localized and delocalized states of small molecules containing 3$d$ transition metals.


We found that the perturbative one-shot $G_{0}W_{0}@\text{PBE}$ scheme, which is the most widely used $GW$ scheme for extended systems, has a couple of problems for finite systems. Fundamentally, $G_{0}W_{0}@\text{PBE}$ underestimates the IE and the $3d$-electron binding energy by $\sim$1~eV and $\sim$2~eV, respectively, which are considerably larger than the widely reported underestimation error of $\sim$0.5~eV. Due to the orbital-character-dependent non-uniform underestimations of $GW$ binding energies, $G_{0}W_{0}@\text{PBE}$ sometimes gives the incorrect orbital order. Practically, $G_{0}W_{0}@\text{PBE}$ suffers from the $GW$ multi-solution issue due to the large distance between QP and gKS-PBE eigenvalues and the complicated pole (peak) structure in the self-energy (the spectral function).


We found that the $G_{0}W_{0}$ starting-point approach, $G_{0}W_{0}@\text{PBE}\alpha$, can improve $G_{0}W_{0}@\text{PBE}$ at the expense of introducing a couple of problems. The $G_{0}W_{0}$ starting-point approach can give good results for the IE and the $3d$-electron binding energy at the same time, and thus, correct the wrong orbital order produced by PBE. Also, the $G_{0}W_{0}$ starting-point approach can mitigate the $GW$ multi-solution issue by reducing the distance between QP and gKS eigenvalues. However, the optimal amount of EXX in the $G_{0}W_{0}$ starting point depends strongly on the amount of 3$d$ character in molecular orbitals, leading to the strong sensitivity of 3$d$-electron binding energy to a change in the EXX amount. Thus, the optimal amount of EXX is strongly system- and property-dependent. More importantly, $G_{0}W_{0}@\text{PBE}\alpha$ suffers from the SCF convergence issue in open-shell systems, which is absent in $G_{0}W_{0}@\text{PBE}$.


We found that the eigenvalue self-consistency approaches, $G_{n}W_{0}@\text{PBE}$ and $G_{n}W_{n}@\text{PBE}$, can improve $G_{0}W_{0}@\text{PBE}$, too. Especially, $G_{n}W_{0}@\text{PBE}$ gives as good results for the IE and the $3d$-electron binding energy as $G_{0}W_{0}@\text{PBE}\alpha$ without suffering from $GW$ multi-solution and SCF convergence issues.



We recommend $G_{n}W_{0}@\text{PBE}$ because of its \emph{practical} advantages: (i) $G_{n}W_{0}@\text{PBE}$ is transferable, because it gives satisfactorily accurate results for both finite and extended systems, for both closed- and open-shell systems, and for both localized and delocalized states, (ii) $G_{n}W_{0}@\text{PBE}$ is predictive, because it does not need any system- and property-dependent parameters, and (iii) $G_{n}W_{0}$ is efficient and easy to use, because it does not require computational and human efforts to address SCF convergence and $GW$ multi-solution issues


We attribute the good performance of $G_{n}W_{0}@\text{PBE}$ to the fortuitous cancellation effect: the overscreening of the Coulomb interaction due to the over-delocalized PBE wavefunction is cancelled by the underscreening due to the neglect of vertex corrections. In other words, for $G_{0}W_{0}$ applied to finite systems, PBE is a ``bad'' starting point in the sense that it causes a large ($\sim$1--2~eV) and orbital-character-dependent underestimation error in electron binding energy, but for $G_{n}W_{0}$ applied to finite and extended systems, PBE is a ``good'' starting point in the sense that it accidentally produces the overscreening just as much as vertex corrections do, which is missing in self-consistent $GW$ schemes.

Our results in this work -- (i) $G_{0}W_{0}@\text{PBE}\alpha$($0.25 \le \alpha \le 0.50$) and $G_{n}W_{0}@\text{PBE}$ give good QP energies for molecular orbitals with both weak and strong 3$d$ character, and (ii) the ev$GW$ starting-point dependency is more related to the orbital character than the self-consistency level -- may seem to disagree with some results in the literature, but this is not the case. The origin of the seeming disagreement is that except for $G_{0}W_{0}@\text{PBE}\alpha$($0.00 \le \alpha \le 0.25$), varying the self-consistency level and the starting point generally makes a small ($\sim$0.1~eV) change in QP energy for HOMO with mainly $sp$ character, which is accidentally comparable to individual or combined errors from multiple sources, such as the incomplete basis set, the linearization method in $G_{0}W_{0}$, and the insufficient number of eigenvalues to update in ev$GW$.




$G_{n}W_{0}@\text{PBE}$ is not a conserving and starting-point-independent $GW$ scheme. It is not the most accurate or efficient $GW$ scheme, either. However, $G_{n}W_{0}@\text{PBE}$ gives satisfactory and reliable results for a wide range of systems, such as solids with strong screening and molecules with weak screening, at moderate computational and minimal human efforts, and thus is ideal for automated mass $GW$ and BSE calculations for high-throughput screening and machine learning. Further studies on the performance of more diverse $GW$ schemes on larger and more complex systems containing a broader range of transition metals are needed to extend the range of applicability of the $GW$ approximation.

\section{Supplementary Material}

See supplementary material for more details, results, and discussion.


%
%

%

\begin{acknowledgments}
This work was supported by the U.S. Department of Energy Grant No. DE-SC0017824. The computation for this work was performed on the high performance computing infrastructure provided by Research Computing Support Services at the University of Missouri-Columbia. This research also used resources of the National Energy Research Scientific Computing Center, a DOE Office of Science User Facility supported by the Office of Science of the U.S. Department of Energy under Contract No. DE-AC02-05CH11231. We also would like to thank Bin Shi and Meisam Rezaei for useful discussions in the earlier stage of this work.
\end{acknowledgments}

\bibliography{ScTiCuZnO}

\end{document}